\documentclass[12pt]{article}
\usepackage[usenames, dvipsnames]{xcolor}
\usepackage{jheppub}
\usepackage{lieart}

\usepackage{tocloft}
\usepackage[normalem]{ulem}
\usepackage[]{todonotes}
\usepackage{verbatim}
\usepackage{mathrsfs}
\usepackage{cleveref}

\usepackage{multirow}

\usepackage{caption}
\usepackage{longtable}
\usepackage{afterpage}
\usepackage{tikz}
\usepackage{setspace,caption}
\usepackage{lipsum}
\usetikzlibrary{matrix,arrows,calc}
\usepackage{bbm}
\usepackage{slashed}
\usepackage{graphicx}
\usepackage{subcaption}
\usepackage{psfrag}
\usepackage{tensor}
\usepackage{fouridx}
\usepackage{enumitem}

\usetikzlibrary{arrows,shapes,snakes,automata,backgrounds,petri}

\usepackage{bm}
\usepackage{mdframed}
\usepackage{multirow}
\usepackage{soul}
\usepackage{multicol}
\usepackage{tikz-cd}
\usepackage{rotating}

\usetikzlibrary{arrows}

  \usepackage{mdframed}
\tikzset{
commutative diagrams/.cd,
arrow style=tikz,
diagrams={>=latex}}

\newcommand{\be}{\begin{equation}}
\newcommand{\ee}{\end{equation}}

\usepackage{amsmath,amsthm}
\usepackage{amssymb}
\usepackage{mathtools}

\usepackage{feynmf}
\usepackage{marvosym}

\usepackage{import}

\usepackage[utf8]{inputenc}
\usepackage{amsmath}
\usepackage{tikz}
\usepackage{extpfeil}
\usepackage{amsfonts}
\usepackage{amssymb}
\usepackage{bm}
\usepackage{graphicx}
\usepackage{array}
\usepackage{lipsum}
\usepackage{float}
\usepackage{multirow}
\usepackage{hhline}
\usepackage{tabularx}
\usepackage{graphicx}
\usepackage{afterpage}
\usepackage{longtable}
\usepackage{epstopdf}
\usepackage{ytableau}
\usetikzlibrary{decorations.markings}

\usepackage[titletoc]{appendix}
\makeatletter
\newtheorem*{rep@theorem}{\rep@title}
\newcommand{\newreptheorem}[2]{%
\newenvironment{rep#1}[1]{%
 \def\rep@title{#2 \ref{##1}}%
 \begin{rep@theorem}}%
 {\end{rep@theorem}}}
\makeatother

\newreptheorem{lemma}{Lemma}

\newreptheorem{conj}{Conjecture}

\theoremstyle{definition}

\theoremstyle{remark}
\newtheorem*{rem*}{Remark}

\newcommand{\jht}[1]{{}}

\newcommand{\ri}{i\,}
\newcommand{\tM}{M_6}

\usepackage{breqn}
\newcommand{\bdm}{\begin{dmath}}
\newcommand{\edm}{\end{dmath}}

\definecolor{cobalt}{RGB}{44, 98, 120}
\definecolor{celadon}{rgb}{0.67, 0.88, 0.69}
\definecolor{dm}{cmyk}{.20, 0, .30, 0}
\definecolor{burgundy}{rgb}{0.5, 0.0, 0.13}
\definecolor{plotBlue}{RGB}{94, 130, 181}

\newcommand*\xoverline[2][0.75]{
    \sbox{\myboxA}{$\m@th#2$}
    \setbox\myboxB\null
    \ht\myboxB=\ht\myboxA
    \dp\myboxB=\dp\myboxA
    \wd\myboxB=#1\wd\myboxA
    \sbox\myboxB{$\m@th\overline{\copy\myboxB}$}
    \setlength\mylenA{\the\wd\myboxA}
    \addtolength\mylenA{-\the\wd\myboxB}
    \ifdim\wd\myboxB<\wd\myboxA
       \rlap{\hskip 0.5\mylenA\usebox\myboxB}{\usebox\myboxA}%
    \else
        \hskip -0.5\mylenA\rlap{\usebox\myboxA}{\hskip 0.5\mylenA\usebox\myboxB}%
    \fi}
\makeatother

\title{\boldmath 
Higher Form and Higher Group Symmetries \\ \vspace{0.5cm} via Mirror Symmetry}

\author{Jiahua Tian$^1$}
\affiliation{$^1$School of Physics and Electronic Science, East China Normal University, \\Shanghai, China, 200241}
\emailAdd{jtian1905@gmail.com}

\author{Xin Wang$^{2,3}$}
\affiliation{$^2$Interdisciplinary Center for Theoretical Study, University of Science and Technology of China, Hefei, Anhui, China, 230026}
\affiliation{$^3$Peng Huanwu Center for Fundamental Theory, \\ Hefei, Anhui, China, 230026}
\emailAdd{wxin@ustc.edu.cn}

\abstract{
In this work we discuss a connection that relates the 1-form and the 2-group symmetries of 5D SCFTs derived from geometric engineering methods to monodromies of the corresponding B-models via mirror symmetry. Viewing defects as branes wrapping relative cycles in a non-compact CY3, we find that the defect groups can be read off from the VEVs of the corresponding line operators at the leading order. Via mirror map, we find that both the 1-form and the 2-group symmetries of the SCFT are related to the monodromy at the large radius point in the B-model. Additionally, we recursively obtain closed-form expressions of instanton expansions of the VEV of Wilson lines of certain 5D theories among which some have not been obtained so far using localization methods. We further conjecture that the 2-group symmetry is given by the Mordell-Weil torsion of the universal special geometry associated to the theory, generalizing the conjecture for rank-1 theories.}

\preprint{USTC-ICTS/PCFT-25-13}
\begin{document}

\maketitle
\flushbottom


\section{Introduction and summary}

Symmetry is a fundamental property of quantum field theory. Recent years have witnessed great progress in the study of \emph{generalized global symmetries} in which the observations made in previous works~\cite{Alford:1990fc, Alford:1991vr, Bucher:1991bc, Alford:1992yx, NUSSINOV2009977, Witten:AdSCFT_TFT, Freed:FluxUncertainty, Pantev:2005zs, Pantev:2005wj, Pantev:2005rh, Hellerman:2006zs, Gukov:2006jk, Aharony:ReadingLines4D, Kapustin:2013uxa, Kapustin:2014gua} were summarized and lifted in the foundational work~\cite{Gaiotto:GenSymm} of this vast subject. Falling under the broad category of generalized symmetries, higher form symmetries, higher group symmetries, and non-invertible symmetries have been extensively studied in the past decade~\cite{Albertini:HigherFormMth, Apruzzi:SymTFT, Apruzzi:GlobalForm_2group, Apruzzi:2group_6D, Apruzzi:Higher_Form_6D, Apruzzi:Holography_1form_Confinement, Apruzzi:NonInvert_Holography, Apruzzi:AspectsSymTFT, Acharya:2023bth, Bashmakov:NonInvClassS, 
vanBeest:SymTFT3D, BenettiGenolini:2020doj, 
Bergman:GenSymmHoloABJM, Bhardwaj:Higher_form_5D6D, Bhardwaj:2Group_S, 
Bhardwaj:NonInvHigherCat, Bhardwaj:GenChargeI, Bhardwaj:GenChargeII, Bhardwaj:1formClassS, Bhardwaj:AnomalyDefect, Bhardwaj:UniversalNonInv, Bhardwaj:UnifyingConstructionNonInv, Bhardwaj:NonInvWeb, Braeger:2024jcj, Choi:NonInv3+1, Cvetic:0form_1form_2group, Cvetic:HigherFormAnomaly, Cvetic:Fluxbrane, 
Cvetic:GensymmGravity, Cvetic:2024dzu, Cvetic:2025kdn, DelZotto:2groupMth, DelZotto:HigherFormAD, DelZotto:HigherFormOrbifold, Etheredge:BraneSymm, GarciaEtxebarria:BraneNonInv, GarciaEtxebarria:Goldstone, Gukov:2020btk, Heckman:Branes_GenSymm, Heckman:2024oot, Hsieh:Inflow, Hubner:GenSymm_EFib, Jia:2025jmn, Kaidi:KW3+1D, Kaidi:SymTFT, Kaidi:NonInvTwist, Lee:MatchingHigher, Liu:2024znj, Morrison:5D_higher_form, Tian:2021cif, Tian:2024dgl}. The interested readers can also read the excellent reviews~\cite{Schafer-Nameki:ICTP, Bhardwaj:lecture, Luo:review, Gomes:review}.

String theory provides not only a means of geometrically engineer quantum field theories, but also a way to understand their generalized global symmetries via string compactification. In this work, we will focus on the study of generalized symmetries of 5D SCFTs, or more precisely its circle reduction which can be constructed via M-theory compactification on the product of a circle and a non-compact Calabi-Yau threefold (CY3)~\cite{Jefferson:Towards5D, Jefferson:Classification5D, Apruzzi:5DSCFT_graphs, Apruzzi:FiberFlavorI, Apruzzi:FiberFlavorII, Apruzzi:5DDecoupleGlue, Closset:Uplane, Closset:5D_Phases_M/IIA, Closset:5D4D, Closset:CBHB0, Closset:CBHBI, Eckhard:Trifectas, Acharya:2021jsp, Tian:2021cif}. A proposal developed in~\cite{GarciaEtxebarria:BraneNonInv, Apruzzi:NonInvert_Holography, Heckman:Branes_GenSymm, Cvetic:Fluxbrane} exemplifies the effectiveness of viewing the charged defects in these quantum field theories as branes wrapping non-compact cycles in the non-compact CY3 in understanding the generalized symmetries. In particular, it was proposed in~\cite{GarciaEtxebarria:BraneNonInv, Heckman:Branes_GenSymm} that the line defects in 5D SCFT can be geometrically engineered as M2-branes wrapping non-compact 2-cycles in a CY3 which can schematically be summarized as:
\begin{equation}
    \text{Line defect on }\ell = \text{M2-brane on }\ell \times \Sigma
\end{equation}
where $\ell$ is a locus in spacetime and $\Sigma$ is a non-compact 2-cycle in CY3 intersecting the boundary of CY3 along a 1-cycle at infinity. These line defects consitute a \emph{defect group}~\cite{DelZotto:2015isa} that encodes the 1-form symmetry of the theory. From now on we will denote by $M_6$ the non-compact CY3 and $\mathcal{T}_{M_6}$ the corresponding 5D quantum field theory.

More precisely, we focus on a 5D $\mathcal{N}=1$ supersymmetric quantum field theory $\mathcal{T}_{M_6}$ with gauge group $G$ on the Euclidean spacetime ${M}^5=\mathbb{R}^4\times S^1$, where $S^1$ is the compactified time direction. On the Coulomb branch of the moduli space, the BPS partition function can be explicitly computed via equivariant localization on the $\Omega$-deformed background \cite{Nekrasov:2002qd}. Such calculations can also be captured by the refined topological string partition function on $M_6$. Now we put a half-BPS, static, infinitely massive quark in the representation $\mathbf{R}$ at the origin of the space, this generates a half-BPS Polyakov loop operator in the representation $\mathbf{R}$
\begin{equation}\label{eq:WilsonOp}
    W_{\mathbf{R}}(\ell) =\mathrm{Tr} \,\mathcal{T} \exp\left(i{\int_{\ell}dt(A_0(t)-\phi(t))}\right),
\end{equation}
which is a special case of the Wilson loop operator wrapping along the compactified Euclidean time direction. Here $\mathcal{T}$ is the time ordering operator, $A_0(t)$ and $\phi(t)$ are the gauge field and scalar field in the vector multiplet that are evaluated at the origin of the space $\mathbb{R}^4$. The scalar field is added to keep half of the supersymmetries.  On the Coulomb branch of the moduli space, the scalar $\phi$ has non-zero expectation values in the Cartan subalgebra of the gauge group $G$, then the gauge group is broken to its Abelian subgroup $U(1)^{r}$ where $r=\mathrm{rank}(G)$ is the rank of the gauge group, which is equal to the 4-th Betti number $b_4$ of the non-compact Calabi-Yau threefold $M_6$. The data of representation is encoded in the electric charges of the Wilson loop particles under each $U(1)$, hence are labeled by the non-negative integers $[q_1,\cdots,q_r]$. Then the Wilson loop operator defined in~(\ref{eq:WilsonOp}) are precisely the one reduced from the M2-brane action on $\Sigma$, where the electric charges $q_i$ are defined via  $q_i=D_i\cdot \Sigma,\, D_i\in H_4(M_6,\mathbb{Z})$, the intersection number of the non-compact curve $\Sigma$ and the compact surfaces $D_i$ in $M_6$.

The Polyakov loop operator is a gauge invariant operator, however, it in general does have charges under the center symmetry $Z(G)$. 
Define $g\in G$ the gauge transformation, which is periodic under the period $\beta$ of $S^1$
\begin{equation}
    g(\vec{x},x_0+\beta)=g(\vec{x},x_0).
\end{equation}
A center symmetry transformation is a gauge equivalence class $h_k\in G$ with the identification $h\sim g h g^{\dag}$ which is periodic up to a center element of $G$. For the case $G=SU(N)$,
\begin{align}
    h_k(\vec{x},x_0+\beta)=e^{\frac{2\pi ik}{N}}h_k(\vec{x},x_0).
\end{align}
The center symmetry acts trivially on any local operators, but it does give phases to the Polyakov operators. In the modern language of symmetries, if the center symmetry is not broken, it is exactly the one-form symmetry of the theory.

It is known that the BPS partition function of the above 5D supersymmetric quantum field theory $\mathcal{T}_{M_6}$ are equal to the partition function of the topological strings on $M_6$. Via mirror symmetry \cite{Candelas:PairCY, Witten:Mirror_TFT, Hosono:MirrorSymmetryHypersurhace, Hosono:MirrorSymmetryCICY, Chiang:LocalMirror}, the calculations can be made on the B-model side of the mirror CY3 $W_6$~\cite{Aganagic:TopologicalVertex, Aganagic:TopologicalString_Int, Klemm:TopoStringAmp, Aganagic:TopoString_ModForm, Huang:TopoString_Modularity, Bouchard:Remodel_Bmodel}. Therefore it is natural to ask if one-form symmetries can be studied from the B-model side of the topological strings?

This question can be answered when we study VEVs of the Wilson loop operators in the B-models. It was conjectured in~\cite{Huang:2022hdo} that the VEVs of the 5D Wilson loop operators are proportional to the complex structure parameters $u_i$ in the B-model of the topological string theory on the mirror manifold $W_6$. This correspondence can be easily verified in the toric CY3 cases where the mirror curve, a section of the B-model geometry, is precisely the Seiberg-Witten curve of $\mathcal{T}_{M_6}$, and the complex structure parameters $u_i$ are equal to the $u$-plane parameters which are the holonomies of the gauge fields. In this paper we show that, as a symmetry, the action of the one-form symmetry at the level of the partition function corresponds to the monodromy transformations for the B-model of topological strings. Furthermore, part of the monodromy actions precisely gives the desired charges of the Wilson loops under one-form symmetries. We present a heuristic explanation from purely quantum field theoretical considerations of this correspondence in Section~\ref{sec:motivation}.

This work is organized as follows. In Section~\ref{sec:motivation} we relate the generalized symmetries of a QFT to certain monodromies in its moduli space via a somewhat heuristic discussion from pure QFT perspective, which will be made concrete and precise in subsequent sections via mirror symmetry. In Section~\ref{sec:WilsonLine_M2brane} we construct Wilson line operators from M2-brane wrapping modes in M-theory compactification then write their VEVs as functions of the K\"ahler moduli parameters of the internal geometry. In Section~\ref{sec:Match_Calc_Bmodel} we will discuss the relation between the 1-form symmetry of the 5D theory which is $\mathcal{T}_{M_6}$ on $S^1$ and the shift of the $B_2$ field in the corresponding IIA picture. In general, for a non-compact CY3, the dimension $b_2$ of the curve classes is larger than the dimension $b_4$ of the compact surface classes, providing additional K\"ahler parameters in the A-model and additional complex structure parameters in the B-model. These additional parameters represent flavors masses and instanton counting parameters on the extended Coulomb branch of $\mathcal{T}_{M_6}$. So in general, the actions of the monodromy group are not always the one-form symmetric actions, but they are the actions of the 2-group symmetry as discussed in Section \ref{sec:Subtlety_Mirror}. We will present our main result in Section~\ref{sec:main_result} where we relate the monodromy of the B-model at the large radius point with the  1-form and the 2-group symmetries of the 5D theory. In Section~\ref{sec:examples_toric} and~\ref{sec:examples} we provide concrete examples to test our statements. As a byproduct, we are able to obtain (order-by-order) closed-form results of the instanton expansions of certain theories via the application of mirror symmetry methods, which have not been calculated before via the traditional localization techniques. In Section~\ref{sec:phys_consequences} we will discuss some of the physical consequence of our calculation in Section~\ref{sec:Defectsfrombranes} and~\ref{sec:partition} and propose in Section~\ref{sec:conjecture} a conjecture relating the topological data of the special geometry of the theory and its 2-group symmetry, generalizing the relevant discussions in~\cite{Closset:Uplane, Cecotti:2024jbt}. In Section~\ref{sec:conclusion} we summarize our results and discuss potential applications and generalizations of the method developed in this work.

\section{Motivation from quantum field theory considerations}\label{sec:motivation}

In this section, we review some general aspects of the action of the symmetry operator upon a defect using the path integral formalism of quantum field theory (see e.g.~\cite{Gomes:review, Bhardwaj:lecture, Schafer-Nameki:ICTP, Jia:2025jmn}). The discussion in this section is intended to motivate the rest of the work. In particular, we aim to illustrate the relationship between certain monodromy actions in the moduli space of a QFT and its generalized symmetries~\footnote{More precisely, the generalized symmetries to be discussed are actually the symmetries at the SCFT point in the moduli space of the QFT, and we will see in Section~\ref{sec:Defectsfrombranes} that the monodromy is the one around the point of large complex structure limit in the moduli space of the mirror geometry.}. We will make the following somewhat heuristic arguments precise in the subsequent sections.

We will focus on abelian higher form symmetries of certain quantum field theory $\mathcal{T}$ in spacetime $M_d$. It is known that the action of a topological symmetry operator $\mathcal{O}$ supported on $\Sigma$ upon a charged defect $W$ supported on $\ell$ is~\cite{Gaiotto:GenSymm}:
\begin{equation}\label{eq:gensymm_action}
    \langle \mathcal{O}(\Sigma) W(\ell) \rangle = e^{i\theta} \langle W(\ell) \rangle
\end{equation}
with a phase factor $e^{i\theta}$. The LHS of the above equation, written in the path integral formalism, is:
\begin{equation}
    \langle \mathcal{O}(\Sigma) W(\ell) \rangle = \int \mathcal{D}\phi\ e^{iS[\phi]}\ \mathcal{O}(\Sigma)\ W(\ell)
\end{equation}
where we denote by $\phi$ the set of local fields in the theory whose action is $S[\phi]$. We absorb the field insertion $\mathcal{O}(\Sigma)$ into a modification of the action $S$ to $S'$ as:
\begin{equation}\label{eq:absorption}
    \int \mathcal{D}\phi\ e^{iS[\phi]}\ \mathcal{O}(\Sigma)\ W(\ell) = \int \mathcal{D}\phi\ e^{iS'[\phi]}\ W(\ell).
\end{equation}
In the above equation, the dependence of $W(\ell)$ on $\phi$ is omitted for simplicity. The absorption in~(\ref{eq:absorption}) can be done, e.g. for an abelian global symmetry with conserved Noether current $j$ in which case $\mathcal{O}(\Sigma) = e^{i \int_\Sigma \star j} = e^{i \int_{M^d} A_{\Sigma}\wedge \star j}$ with closed $\Sigma$ and $A_{\Sigma}$ is the Poincar\'e dual of $\Sigma$ in $M^d$. Thus we arrive at the new action $S' = S + \int_{M^d} A_\Sigma\wedge \star j$. Suppose that there exists a change of variables $\phi\rightarrow\phi'$ such that the functional form of $S$ is recovered while the dependence of $W(\ell)$ on $\phi$ remains unchanged, we have:
\begin{equation}\label{eq:Recover_Shift}
    \int \mathcal{D}\phi\ e^{iS'[\phi]}\ W(\ell) = \int \mathcal{D}\phi'\ e^{iS[\phi']}\ W(\ell).
\end{equation}
To see the existence of such a change of variables, let us look at $U(1)$ gauge theory with electric 1-form symmetry as a concrete example. In this case the topological operator is~\cite{Bhardwaj:lecture}:
\begin{equation}
    \mathcal{O}(\Sigma) = \exp\left( i\alpha \int_\Sigma \star F \right),\ g = e^{i\alpha}
\end{equation}
for a $(d-2)$-dimensional closed surface $\Sigma$, which can be further rewritten as:
\begin{equation}
    \mathcal{O}(\Sigma) = \exp\left( i\alpha \int_{M_d} \delta_\Sigma\wedge \star F \right)
\end{equation}
for a 2-form $\delta_\Sigma$ with delta function support on $\Sigma$. To obtain the absorption~(\ref{eq:absorption}), we consider the shift $A\rightarrow A - \frac{1}{2} \frac{\alpha}{2\pi} d\varphi$, where $\varphi$ parametrizes a small circle linking $\Sigma$ in $M_d$ and $d(d\varphi) = 2\pi \delta_\Sigma$~\cite{Beasley:2009mb}. For such shift, we have:
\begin{equation}
    S[A] \rightarrow S'[A] = S[A] - \alpha \int_{M_d} \delta_\Sigma \wedge \star F + \frac{\alpha^2}{4} \int_{M_d} \delta_\Sigma\wedge \star \delta_\Sigma
\end{equation}
where the second term is exactly what is needed to absorb the insertion of $\mathcal{O}(\Sigma)$, and the last term represents a formally divergent ``self energy'' of $\Sigma$ to be regularized by adding a counter term~\footnote{In 3D for Chern-Simons theory this would be the self-linking of a knot, which can be regularized by the point-splitting regularization~\cite{Witten:1988hf}. Here in principle we care only about the correlator $\langle \mathcal{O}_\Sigma W(\ell) \rangle$ and the self-linking, or self energy, of the operators are unimportant for our purpose.}. The shift $A\rightarrow A - \frac{1}{2} \frac{\alpha}{2\pi} d\varphi$, however, does not change the functional dependence of $W(\ell) = e^{i\oint_\ell A}$ on $A$ at all since by definition $\ell \cap \Sigma = \emptyset$ hence any term proprotional to $\delta_\Sigma$ integrates to zero over $\ell$. Therefore, in this concrete example, apart from a divergent term to be regularized, the shift $A\rightarrow A - \frac{1}{2} \frac{\alpha}{2\pi} d\varphi$ is the desired field redefinition for~(\ref{eq:Recover_Shift}) to hold.

\begin{figure}[h]
    \centering
    \includegraphics[width=0.2\linewidth]{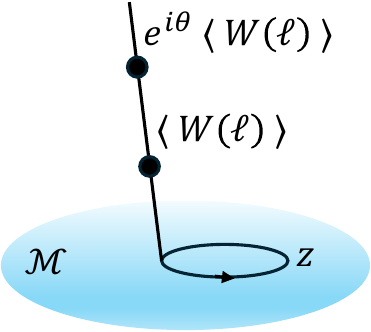}
    \caption{$\langle W(\ell) \rangle$ picks up a phase when $z$ loops in $\mathcal{M}$.}
    \label{fig:moduli_z}
\end{figure}

Looking back at the RHS of~(\ref{eq:gensymm_action}), it can be written as:
\begin{equation}
    e^{i\theta} \int\mathcal{D}\phi\ e^{iS[\phi]}\ W(\ell).
\end{equation}
Therefore, we have, at least schematically:
\begin{equation}\label{eq:loop_z}
    \int \mathcal{D}\phi'\ e^{iS[\phi']}\ W(\ell) = e^{i\theta} \int\mathcal{D}\phi\ e^{iS[\phi]}\ W(\ell).
\end{equation}
Now we make another key assumption that $\phi$ collectively depends on a moduli $z$ living in certain moduli space $\mathcal{M}$. Since we have started with a theory whose action is $S$ and ended on the same action $S$, it is natural to view the change of variables $\phi\rightarrow\phi'$ as $z$ traversing a loop in $\mathcal{M}$. Therefore,~(\ref{eq:loop_z}) can be understood as:
\begin{equation}
    \langle W(\ell) \rangle_{e^{2\pi i}z} = e^{i\theta} \langle W(\ell) \rangle_z,
\end{equation}
i.e. $\langle W(\ell) \rangle$ picks up a phase after $z$ has traversed a loop in $\mathcal{M}$ ~(see Figure~\ref{fig:moduli_z}).

Based on the above discussions, one can view $\langle W(\ell) \rangle$ as living in a line bundle over $\mathcal{M}$ whose monodromy is determined by the phase factor $e^{i\theta}$, of which the simplest case is $\theta = 2\pi/n$ for a higher form $\mathbb{Z}_n$-symmetry. A toy model exhibiting such an $e^{2\pi i/n}$ (or equivalently $e^{-2\pi i/n}$) monodromy is the following multi-valued complex function~\footnote{We choose to work with the negative power monodromy in order to match the precise calculation in Section~\ref{sec:partition}.}:
\begin{equation}\label{eq:toy_model}
    u = z^{-1/n}.
\end{equation}
Writing $z$ as a function of $u$, we have:
\begin{equation}\label{eq:rough_relation}
    z = \frac{1}{u^n}
\end{equation}
and it is trivial to see that the transformations (mod $\mathbb{Z}$) of $u$ that keep $z$ invariant is generated by $u\rightarrow e^{2\pi i/n} u$, which is isomorphic to the higher form symmetry group $\mathbb{Z}_n$. In other words, after writing $z$ as a function of $u$, one can identify the group of transformations of $u$ that keep $z$ invariant with the higher form symmetry group of $\mathcal{T}$ if $u$ is identified with $\langle W(\ell) \rangle$.

The lesson we learn from the discussions in this section is that the higher form symmetry of a QFT, which traditionally is viewed as a property of the QFT itself by manipulating its local fields as in~(\ref{eq:Recover_Shift}) and~(\ref{eq:loop_z}), can actually be related to monodromies in its moduli space. We will see in the following sections that the above heuristic arguments from a purely quantum-field-theoretical point of view can be made very precise. Indeed, we will show that $z$ is the complex structure moduli of the mirror of the Calabi-Yau threefold on which we compactify M-theory to geometrically engineer theory $\mathcal{T}$. In particular, the monodromy exhibited by the toy model~(\ref{eq:toy_model}) will be made precise in Section~\ref{sec:Defectsfrombranes} and~\ref{sec:partition} using mirror symmetry and indeed we will find that $\langle W(\ell) \rangle = z^{-1/n}$ at the leading order of the mirror map.

\section{Defect lines from geometric engineering and the mirror}\label{sec:Defectsfrombranes}

In this Section we discuss some basic properties of Wilson lines in 5D theory from various angles. In particular, we focus on the geometrically engineered 5D theory via M-theory on a non-compact CY3 $M_6$, which we denote by $\mathcal{T}_{M_6}$, and study the Wilson lines that are obtained from M2-brane wrapping non-compact cycles that are extended charged objects under the 1-form symmetry of the 5D theory~\cite{Gaiotto:2014ina}. We will further reduce $\mathcal{T}_{M_6}$ on $S^1$ and discuss the relevant properties of the resulting 4D KK theory.

In Section~\ref{sec:WilsonLine_M2brane} we construct the line defect $W(\ell)$ of $\mathcal{T}_{M_6}$ on $S^1$ from M2-brane wrapping a non-compact 2-cycle in $M_6$ and the extra $S^1$ in the spirit of~\cite{Heckman:Branes_GenSymm} and calculate $\langle W(\ell) \rangle$ as a function of the K\"ahler parameters of the compactification. In Section~\ref{sec:Match_Calc_Bmodel} we discuss the relation between 1-form symmetry and $B_2$ field shift in the IIA compactification corresponding to $\mathcal{T}_{M_6}$ on $S^1$. In Section~\ref{sec:Subtlety_Mirror} we point out an important subtlety in the calculation of $\langle W(\ell) \rangle$ and its relation to the 2-group symmetry of the theory.

\subsection{Defect lines from wrapped M2-brane branes}\label{sec:WilsonLine_M2brane}

In M-theory compactification each harmonic 2-form $\omega_i \in H^2_{\text{cpt}}({\tM}, \mathbb{Q}), i=1,\cdots,r$ for a smooth CY3 ${\tM}$ leads to a 5D gauge field via the 3-form gauge potential 
$C_3 = \sum_i A_i\wedge \omega_i$. Here $r=\mathrm{rank}(H^2_{\text{cpt}}({\tM}, \mathbb{Q}))$ is the rank for the second de Rham cohomology with compact support. In other words, each $U(1)$ gauge field $A_i$ corresponds to a compact divisor $D_i\in H_4(M_6,\mathbb{Z})$ which is dual to certain $\omega_i$.  The gauge coupling of $A_i$ is set by $g_i^2 \sim 1/\text{vol}(D_i)$. The theory $\mathcal{T}_{{\tM}}$ generically has a $U(1)^r$ gauge group which enhances to non-abelian gauge group when ${\tM}$ becomes singular. The decomposition of $C_3$ is subtle for singular $M_6$, nevertheless we will see such subtlety does not play any substantial role in our analysis.

By shrinking the size of compact surfaces in $M_6$ to zero, the non-compact $M_6$ is a cone over its boundary $\partial M_6$ with metric $ds_{M_6}^2 = dr^2 + r^2 ds_{\partial M_6}^2$ with $r \in [0,\infty)$. To regularize the result, which will be discussed in a moment, we cut off $r$ at large $R$ and later send $R$ to infinity. This cut-off in the cone direction breaks the Ricci-flatness of $M_6$ which is recovered when $R\rightarrow \infty$. For finite $R$, $M_6$ becomes a manifold with boundary at $r = R$.

\begin{figure}[h]
    \centering
    \includegraphics[width=0.7\linewidth]{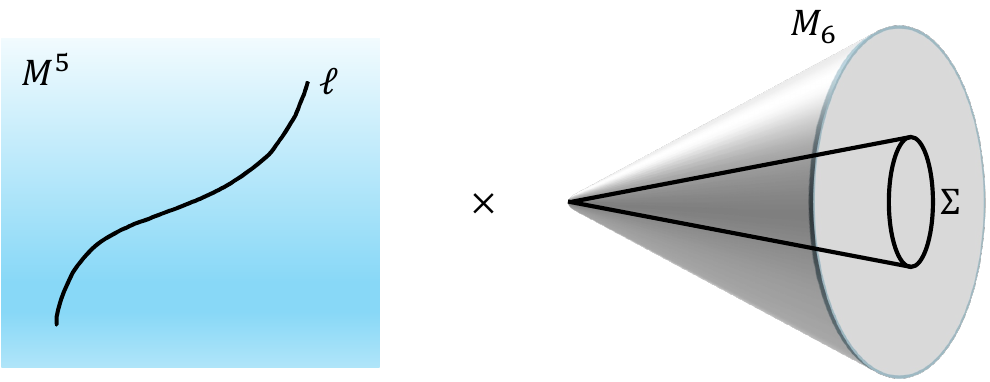}
    \caption{A line defect $\ell\in M^5$ as an M2-brane wrapping $\ell\times\Sigma$.}
    \label{fig:brane_wrapping}
\end{figure}

A line defect in $\mathcal{T}_{M_6}$ in spacetime $M^5$ is constructed by M2-brane wrapping $\ell\times \Sigma$ for a world line $\ell \subset M^5$ and a relative 2-cycle $\Sigma\subset M_6$  (see Figure~\ref{fig:brane_wrapping}) modulo charge screening by compact 2-cycles, i.e. we have~\cite{Morrison:5D_higher_form, Heckman:Branes_GenSymm}:
\begin{equation}
    \Sigma \in H_2(M_6, \partial M_6)/H_2(M_6),
\end{equation}
where in here and in the following the (relative) homology groups are taken with coefficient group to be $\mathbb{Z}$ unless otherwise specified.
In this work we will always assume $M^5 = \mathbb{R}^4\times S^1$ and set $\ell=S^1$. In this sense we are actually studying the $S^1$-compactified 5D theories, or equivalently, the 4D $\mathcal{N}=2$ Kaluza-Klein type (KK) theories. 
The classical action of such an M2-brane is $\int_{\ell\times\Sigma} C_3$ plus a volume form, the integration of $C_3$ provides the antisymmetric $B_2$ field or in short $B$-field in the resulting IIA theory while the integration over the volume form gives the volume of $\Sigma$, so the total contribution gives provides a complexified K\"ahler parameter $t_{\Sigma}$. The VEV of the Wilson loop operator gives the number of such M2-branes on $\Sigma^{\prime}\in H_2(M_6, \partial M_6;\mathbb{Z})$ which has degree one component of $\Sigma$, therefore we have:
\begin{equation}\label{eq:WilsonVEV}
    \langle W(\ell) \rangle = e^{2\pi i t_{\Sigma}}\left(1+\mathcal{O}(e^{2\pi i t_i})\right),
\end{equation}
where $t_i$ are the complexified K\"ahler parameters of $M_6$. The leading coefficient in \eqref{eq:WilsonVEV} is always one because in the limit $R\rightarrow \infty$, only single non-dynamic M2-brane contributes.

In the limit $t_{\Sigma}\rightarrow \infty$, the massive quark is decoupled from the system, thus
the VEV of the Wilson loop (\ref{eq:WilsonVEV}) is zero. To make sense of the calculation, the concept {\emph{compact representative}} was introduced in \cite{Bhardwaj:Higher_form_5D6D, Hubner:GenSymm_EFib}. Recall that the 11D three-form gauge field $C_3$ has a decomposition $C_3 = A \wedge \omega^\Sigma$ on $\Sigma$, for $\omega^\Sigma \in H_{\text{cpt}}^2(M_6,\partial M_6)$ such that the pairing $\kappa_R := \int_\Sigma\omega^\Sigma$ is non-trivial, which is essentially the volume of $\Sigma$ proportional to the cut-off $R$~\footnote{Actually, as a cone over a 1-cycle $\gamma\subset \partial M_6$, $\Sigma$ depends on $R^d$ for certain $d$. But this power is not important for our purpose.}. If we take the limit $R \rightarrow \infty$, the volume of $\Sigma$, which is proportional to the mass of the massive quark, goes to infinity. Subtracting this infinity contribution we find a finite result for the VEV of the Wilson loop operator, which can be computed by introducing the \emph{compact representative} 
\begin{align}\label{eq:find_cpt_reps}
\Sigma_{c,k} = \sum_i \mathfrak{c}^{(k)}_i \Sigma_i,\quad k=1,\cdots,b_4
\end{align}
for $ \Sigma_i \in H_2({\tM},\mathbb{Z})$ and $\mathfrak{c}^{(k)}_i\in \mathbb{Q}$ such that
\begin{align}
    D_i\cdot \Sigma_{c,k}=\delta_{ik}.
\end{align}
Define the gauge charge $q_i=D_i\cdot \Sigma$, then the compact representative $\Sigma_c$ of $\Sigma$ is written as
\begin{align}
    \Sigma_c=\sum_{k=1}^{b_4}q_k \Sigma_{c,k}.
\end{align}
However, the coefficients $\mathfrak{c}^{(k)}_i$ are in general not unique, this is because $b_2\geq b_4$ for a non-compact CY3 so that~(\ref{eq:find_cpt_reps}) is underdetermined. This ambiguity can be fixed if we further choose $f=b_2-b_4$ non-compact surfaces and require that $\Sigma_c$ does not intersect with any of these non-compact surfaces. In the low energy gauge theory phase of $\mathcal{T}_{M_6}$, the choice of non-compact surfaces fixes $f=b_2-b_4$ additional mass parameters on the extended Coulomb branch where $f$ turns out to be the rank for the flavor group. The volume of $\Sigma_c$'s chosen in this way depends only on the Coulomb branch parameters while not on the mass parameters at all.

For pure Yang-Mills theory without any hypermultiplets, the compact representative can be expanded in terms with the curve classes $\Sigma_j^{G}$ that are related to the simple roots of $G$, since $D_i\cdot \Sigma_j^{G}=-C_{ij}$, where $C_{ij}$ is the Cartan matrix defined from the inner product of coroots and roots, we have:
\begin{align}
    \Sigma_{c,k}=-\sum_{j=1}^{b_4}(C^{-1})_{kj}\Sigma_j^{G}.
\end{align}

For simplicity we assume for now that $G = SU(N)$ and only consider the Wilson loop in the fundamental representation. In this case we have $\Sigma_c=-\frac{1}{N}\sum_{j=1}^{N-1} (N-j)\Sigma_{j}^{G}$:
\begin{equation}\label{eq:SUn_Wilson}
    \langle W(\ell) \rangle = e^{-\frac{2\pi i}{N} \sum_{j=1}^{N-1}(N-j)t_j}\left(1+\mathcal{O}(e^{2\pi i t_j})\right).
\end{equation}
If we are only interested in the symmetry charges, we note that the calculations leading to the crucial $1/N$ factor in the $G = SU(N)$ case in~(\ref{eq:SUn_Wilson}) is equivalent to the method used in~\cite{Morrison:5D_higher_form, Bhardwaj:Higher_form_5D6D, Hubner:GenSymm_EFib} to obtain the center divisor as a linear combination of the compact divisors with rational coefficients. We will give a derivation using that method in Appendix~\ref{app:SNF_approach} to show the equivalence.

\subsection{1-form symmetry from $B$-field shift}\label{sec:Match_Calc_Bmodel}

In this section, we discuss how the 5D 1-form symmetry transformation arises from M-theory compactification and then connect it to the topological string theory.

For M-theory compactification on $S^1$, the antisymmetric $B_2$ field in the resulting IIA theory comes from the 3-form gauge field $C_3$ on $S^1$. The integral of the $B_2$ field over a close 2-cycle counts the winding numbers over the circle so the action does change under shift by $2\pi i \,n$, $n\in \mathbb{Z}$. The shift of the $B$-field by a $2\pi i $-phase corresponds to a shift of the complexified K\"ahler parameter:
\begin{align}\label{eq:tshift}
    t_i \rightarrow t_i+1
\end{align}
which is a symmetry of the topological string partition function at large volume. However, this transformation gives a phase to the VEV of Wilson loops due to the expansion~\eqref{eq:SUn_Wilson} so it is natural to expect that it corresponds to the 1-form symmetry of the 5D SQFT.

In the context of local mirror symmetry \cite{ Chiang:LocalMirror,Aganagic:2006wq}, the shift \eqref{eq:tshift} corresponds to part of the monodromy transformation for the B-model periods which shifts the A-periods by constant period 1.
In the B-model of the mirror manifold $W_6$, the partition function is computed from the periods $\Pi(z)=\{\Pi_0(z),\Pi_{\text{A},i}(z),\Pi_{\text{B},i}(z)\}$ of the geometry, which depends on the complex structure moduli $z_i$. In particular, for local CY3, $\Pi_0(z)=1$ is a constant period. Around the large volume point $z\sim 0$, the A-periods define the mirror maps
\begin{align}\label{eq:mirrormap}
    t_i:=\Pi_{\text{A},i}=\frac{1}{2\pi i } (\log(z_i)+\mathcal{O}(z)),\quad i=1,\cdots, b_2,
\end{align}
where $b_2 $ is the second Betti number for $M_6$. So the shifts of $t_i$ as well as other monodromy transformations for the B-periods are generated by
\begin{align}
    z_j\mapsto e^{2\pi i} z_j
\end{align}
in the mirror.

Combining~(\ref{eq:SUn_Wilson}) for $G = SU(N)$ and the mirror map~(\ref{eq:mirrormap}), at leading order we have:
\begin{equation}\label{eq:W_as_z}
    \langle W(\ell) \rangle \sim z^{-\frac{1}{N}}
\end{equation}
where $z$ is the mirror of $t$ which is the K\"ahler parameter of the compact representative corresponding to the generator of $Z(SU(N)) \cong \mathbb{Z}_N$. It is immediate to see that the form of $\langle W \rangle$ matches our toy model~(\ref{eq:toy_model}), therefore by our heuristic argument in Section~\ref{sec:motivation} the theory has $\mathbb{Z}_N$ 1-form symmetry which is indeed true for pure $SU(N)$ theory. We conclude that the application of mirror map upon the VEV of Wilson loop backs up our heuristic derivation in~\ref{sec:motivation}. We will sharpen this observation further in Section~\ref{sec:partition} with more detailed discussions and concrete examples and see it encodes not only the 1-form symmetry but also the 2-group symmetry of $\mathcal{T}_{M_6}$.

\subsection{A subtlety that signals 2-group symmetry}\label{sec:Subtlety_Mirror}

Before heading into concrete examples, the reader may have already noticed a subtlety from the discussion in the last subsection, that is the shift~\eqref{eq:tshift} may not always come from a shift of Coulomb branch parameters, but can instead be from a shift of mass parameters which do not correspond to an element of the genuine one-form symmetry group $\Gamma^{(1)}$. 
In fact in this section we will argue they constitute a group $\mathcal{E}$ that is larger than $\Gamma^{(1)}$ of $\mathcal{T}_{M_6}$. In this section for simplicity we assume $M_6$ is an elliptically fibered CY3 while it does not have to be so for general M-theory compactification.

Let us recall that $\Gamma^{(1)}$ can be embedded in the following short exact sequence when $\mathcal{T}_{M_6}$ exhibits \emph{2-group symmetry}~\cite{Bhardwaj:2Group_S, LeeKantaroTachikawa:2021crt}:
\begin{equation}\label{eq:Extend_LineDefects}
    0 \rightarrow \mathcal{F} \rightarrow \mathcal{E} \rightarrow \Gamma^{(1)} \rightarrow 0
\end{equation}
where $\mathcal{E}$ is the group of ``naive'' line defects where their endings on flavor branes are neglected~\cite{Cvetic:0form_1form_2group}.  
Equivalently, $\mathcal{E}$ is the group of line defects one would have obtained if all flavor brane locus were made compact.

\begin{figure}[h]
    \centering
    \includegraphics[width=0.6\textwidth]{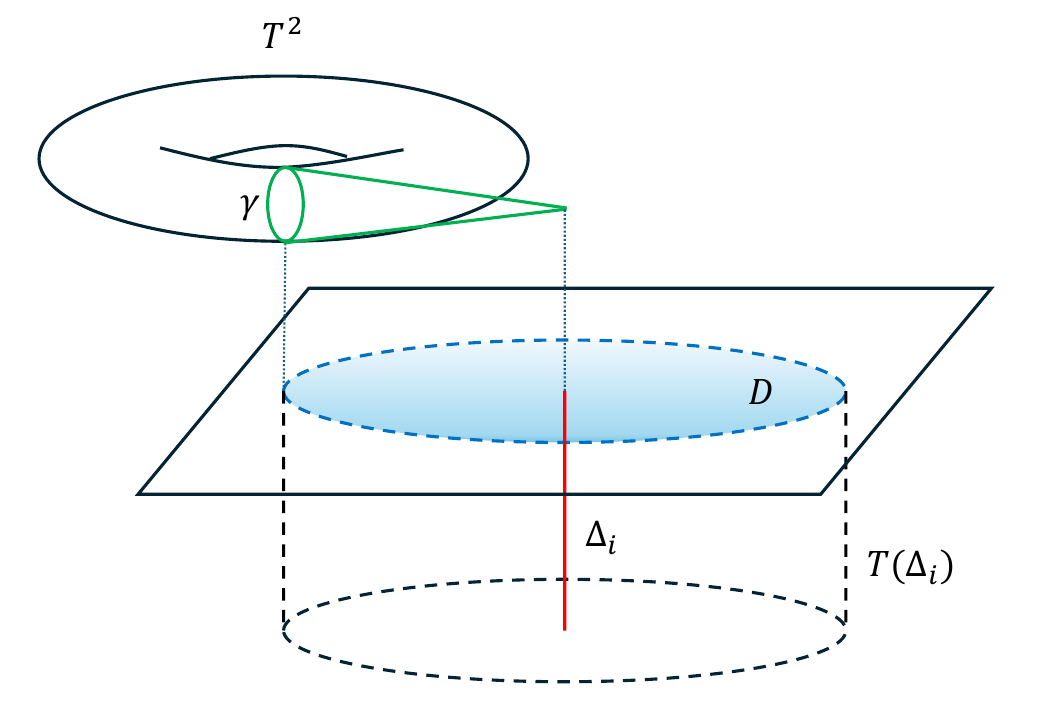}
    \caption{The geometry of $\partial M_6^\circ \cap T(\Delta_i)$ and the flavor Wilson lines in it. A flavor Wilson line is given by an M2-brane wrapping the cone over $\gamma$.}
    \label{fig:FlavorWilson}
\end{figure}

For instance, let us consider a generic 5D KK theory arises from circle compactification of a 6D SCFT, where the 6D theory is obtained from the F-theory compactified on the elliptically fibered Calabi-Yau threefold $M_6$ over a non-compact base. Following the discussion in~\cite{Cvetic:0form_1form_2group}, let $\Delta = \cup_i \Delta_i \subset M_6$ where $\Delta_i$ is the $F_i$-flavor brane locus and we study the geometry of $\partial M_6^\circ \cap T(\Delta_i)$ illustrated in Figure~\ref{fig:FlavorWilson} where $\partial M_6^\circ := \partial M_6\backslash \Delta$ and $T(\Delta_i)$ is a tubular neighborhood of $\Delta_i$.
The region $\partial M_6^\circ \cap T(\Delta_i)$ can be viewed as an elliptic fibration over a punctured disk $D\backslash \{0\}$. The elliptic fibration extended from $D\backslash \{0\}$ to $D$ is an ADE singularity determined by $F_i$. The flavor Wilson lines of $F_i$ are given by M2-brane wrapping $\text{Cone}(\gamma)\times \ell$ for $\gamma\in \text{Tor}H_1(\partial M_6^\circ \cap T(\Delta_i))$. As is now clear from Figure~\ref{fig:FlavorWilson}, we have:
\begin{equation}
    \text{Tor}H_1(\partial M_6^\circ \cap T(\Delta_i)) \cong \text{Tor}H_1(S_{F_i}^\circ)
\end{equation}
where $S_{F_i}^\circ$ is the elliptic fibration $E\hookrightarrow S_{F_i}^\circ \rightarrow D\backslash\{0\}$. It is shown in~\cite{Hubner:GenSymm_EFib, Heckman:Branes_GenSymm} that $\text{Tor}H_1(S_{F_i}^\circ) \cong \text{coker}(\rho_i - 1)$ where $\rho_i$ is monodromy associated to the $F_i$-flavor brane which in the dual F-theory picture becomes 7-branes on $\Delta_i$. The torsion group $\text{coker}(\rho_i - 1)$ is exactly what one gets in 8D F-theory compactification when the 7-branes are gauge branes hence the 8D theory is a pure gauge theory~\cite{Gaiotto:GenSymm, Cvetic:OneLoop}. Via M/F-theory duality this is clearly also true in 7D when $\Delta_i$ can be viewed as a compact cycle, hence we conclude that the flavor Wilson lines become genuine line defects charged under $Z(F_i)$ when the corresponding $\Delta_i$ is compact. This is the reason why $\mathcal{E}$ can be viewed as the group of line defects if all the cycles supporting flavor branes were compact as mentioned in the last paragraph.

Having explained the extension of genuine line defect group $\Gamma^{(1)}$ to $\mathcal{E}$~(\ref{eq:Extend_LineDefects}), we now look back at the descriptions in Section~\ref{sec:Match_Calc_Bmodel}. If the shift \eqref{eq:tshift} only shift the K\"ahler parameters that are related to the Coulomb parameters, it will generate the genuine a one-form symmetry transformation in $\Gamma^{(1)}$. But since the shift may arise from a shift of a mass parameter, what are really generated by \eqref{eq:tshift} are the ``1-form symmetry'' of the ``naive'' line defects (i.e. if we simply ignore the screening of the line defects by the charged matters) in $\mathcal{E}$. Clearly, since any screening can only reduce $\mathcal{E}$ to its subgroup, we have $\Gamma^{(1)} \subset \mathcal{E}$. In the next section we will compute $\mathcal{E}$ and $\Gamma^{(1)}$ in~(\ref{eq:Extend_LineDefects}) concretely applying~(\ref{eq:SUn_Wilson}) and~(\ref{eq:mirrormap}). In particular, we will illustrate the method to distinguish $\Gamma^{(1)}$ from $\mathcal{E}$ and see that this distinction encodes the 2-group symmetry of $\mathcal{T}_{M_6}$.

\section{1-form and 2-group symmetries and B-models}\label{sec:partition}

In this section, we provide a detailed discussion on how to calculate the VEVs for 5D Wilson loops in the topological string B-model, as well as the method for determining 1-form and 2-group symmetries in~(\ref{eq:Extend_LineDefects}). We refer the reader to \cite{Clader:2018yyu} for a review on the topological string B-models. In section~\ref{sec:main_result} we present our main statement that the 1-form and the 2-group symmetries of the 5D theory $\mathcal{T}_{M_6}$ can be identified with certain reparametrization groups of the mirror geometry. In Section~\ref{sec:examples_toric} and~\ref{sec:examples} we present concrete examples to illustrate and test our main claim.

\subsection{1-form and 2-group symmetries and the reparametrization group of partition function}\label{sec:main_result}

Consider the topological string A-model on $M_6$, whose partition function calculates the BPS partition function of $\mathcal{T}_{M_6}$ on the Coulomb branch. Via mirror symmetry, the B-model calculations give the same partition function after a change of variables \eqref{eq:mirrormap}. In the B-model of the mirror manifold $W_6$, the geometry is characterize by a sequence of complex structure parameters $a_i$. In the toric CY3 cases, these parameters can be viewed as the the homogeneous coordinates of the toric CY3 varieties. However, they are not all independent. Under toric actions, some of the parameters $a_i$ can be set to one, leaving independent parameters $u_k$ and $m_k$, which are related to compact and non-compact surfaces in the A-model. 
One can also define the invariant parameters under the toric actions as \footnote{Here the parameters $z_i$ are precisely what we have used in Section \ref{sec:Match_Calc_Bmodel}}
\begin{align}\label{eq:z_um}
    z_i=\prod u_k^{Q^{G}_{ki}}\prod m_l^{Q^F_{li}},
\end{align}
where $Q^{G}_{ki},Q^{F}_{li}$ are integers calculated from the intersection numbers between the $k$-th compact surface or the $l$-th non-compact surface and the $i$-th curve.  \eqref{eq:z_um} holds even if the non-compact Calabi-Yau threefold is described by a hypersurface in a toric variety, in this case, the B-model geometry is a hypersurface in a (weighted) projected space, described by 
\begin{align}
    \mathcal{H}(u,m;Y)=0,\quad Y=(Y_1,\cdots,Y_4),
\end{align}
where $Y_i$ are the inhomogeneous coordinates of the weighted projected space. Consider the transformation on the moduli space parameters:
\begin{equation}\label{eq:trans}
    u_i \mapsto e^{2\pi \ri \lambda_i} u_i,\quad i=1,\cdots,b_4;\quad\quad m_j \mapsto e^{2\pi \ri \kappa_j} m_j, \quad j=1,\cdots,b_2-b_4,
\end{equation}
such that the B-model geometry satisfies
\begin{align}\label{eq:cond0}
     \mathcal{H}(e^{2\pi \ri \lambda}u,e^{2\pi \ri \kappa}m;Y)=e^{2\pi \ri \mu}\mathcal{H}(u,m;e^{2\pi \ri \beta}Y),
\end{align}
then the vanishing loci of $\mathcal{H}(e^{2\pi \ri \lambda}u,e^{2\pi \ri \kappa}m;Y)$ and $\mathcal{H}(u,m;Y)$ are the same hypersurface and they describe the same physical system. See~\cite{Lerche:1989cs,Lerche:1991wm} for related discussions.

In general, the transformation \eqref{eq:trans} that satisfies the condition \eqref{eq:cond0} generates the action of the monodromy transformation on the B-model periods $\Pi(z)$ at the large radius point. It can alternatively be expressed as
\begin{align}\label{eq:zshift}
    z_j\mapsto e^{2\pi i}z_j,
\end{align}
where $z_j$ are invariant coordinates in \eqref{eq:z_um}. Under mirror symmetry, the shift \eqref{eq:zshift} corresponds to the shift of the complexified K\"ahler parameters $t_j\mapsto t_j +1$ which is equivalent to the shift of the $B$-fields.
The consistency of \eqref{eq:trans} and \eqref{eq:zshift} imposes the condition:
\begin{equation}\label{eq:trans2}
    \sum_{k}Q_{ik}^{G}\lambda_k+\sum_l Q_{il}^{F}\kappa_l=0\mod 1, \quad \forall i\,.
\end{equation}
The transformations given by~(\ref{eq:trans}) and \eqref{eq:trans2} form a group denoted by $\mathcal{G}_{um}$ and the subgroup obtained by setting $\kappa_l=0$ is denoted by $\mathcal{G}_u$. These two groups are the key objects that will later be identified with $\mathcal{E}$ and $\Gamma^{(1)}$, respectively.

It is not hard to see that the groups $\mathcal{G}_{um}$ and $\mathcal{G}_{u}$ can be computed via Smith decomposition of the charge matrices $Q_{um,b_2\times b_2}=\{Q^{G},Q^{F}\}$ and  $Q_{u,b_4\times b_2}=Q^{G}$ respectively, where the second subscripts are the dimensions of the charge matrices.
Suppose $\mathcal{M}$ is a $a\times b$ full rank matrix, with $a\leq b$, then the Smith normal form of $\mathcal{M}$ is determined by unimodular matrices $U_{a\times a}$ and $V_{b\times b}$ via
\begin{equation}
    \text{SNF}(Q) = UQV = \begin{pmatrix}
        \alpha_{1} & 0 & \cdots & 0 & \cdots &0 \\
        0 & \alpha_{2} & \cdots & 0 & \cdots & 0 \\
        \vdots & & \ddots & \vdots & & \vdots \\
        0 & \cdots &  & \alpha_{a} & \cdots & 0
    \end{pmatrix},\quad \alpha_i \in \mathbb{Z}_+\, .
\end{equation}
For instance, the Smith normal form for $Q=Q_{u}$ suggests a new set of variables
\begin{align}
    z_{(i)}=\prod_{j=1}^{b_2}z_j^{V_{ji}},\quad u_{(j)}=\prod_{j=1}^{b_4}u_j^{U^{-1}_{ij}},
\end{align}
such that
\begin{align}\label{eq:diagnal_ziui}
    z_{(i)}=u_{(i)}^{\alpha_i}\, m_l^{\sum_{j=1}^{b_2}Q_{lj}^FV_{ji}}.
\end{align}
From~\eqref{eq:diagnal_ziui}, one can determine $\mathcal{G}_{u}=\prod_{i=1}^{b_4}\mathbb{Z}_{\alpha_i}\cong\Gamma^{(1)}$, which is exactly the one-form symmetry derived in~\cite{Morrison:2020ool}. The same method applies to $Q_{um}$ determines $\mathcal{G}_{um}$, where we can further recognize  $\mathcal{G}_{um}\cong\mathcal{E}$ according to~\cite{Apruzzi:2021vcu}. Then 2-group symmetry can also be determined via the short exact sequence
\begin{equation}\label{eq:main_sequence}
    0 \rightarrow \mathcal{F} \rightarrow \mathcal{G}_{um} \rightarrow \mathcal{G}_{u} \rightarrow 0\,.
\end{equation}
By comparing~(\ref{eq:diagnal_ziui}) with~(\ref{eq:W_as_z}), it is easy to see that at leading order we have $u_{(i)} = \langle W_{\mathbf{R}_{(i)}}(\ell) \rangle$ where $\mathbf{R}_{(i)}$ is the the $i^{\text{th}}$ fundamental representation of the gauge group (assuming there is a single gauge group with semi-simple algebra, see~\cite{Huang:2022hdo} for a related conjecture) and~(\ref{eq:W_as_z}) is for $\mathbf{N}$ of $SU(N)$. This presciption can be easily generalized to the cases with several gauge groups.

In a concrete example 5D pure $SU(2)_0$ theory that we will compute in Section~\ref{sec:SU2_0}, the partition function is a function of $z_1 = u^{-2},z_2=mu^{-2}$ and $u$ admits a change in phase by $e^{\pi i}$ under which $z_1,z_2$, hence the partition function is invariant. Hence in this case there is clearly a $\mathbb{Z}_2$ symmetry in the theory. However, this $\mathbb{Z}_2$ symmetry action changes the phase of the Wilson loop VEV by $e^{\pi i}$, providing the correct charge of the fundamental Wilson loop under one-form symmetry.
For the $SU(2)_{\pi}$ theory, the partition function $\mathcal{Z}$ is a function of $z_1 = u^{-2}$, $z_2 = \frac{m}{u}$, that there is no $\mathbb{Z}_2$ symmetry if $\kappa=0$. However, there is still a $\mathbb{Z}_2$ global transformation $(u,m)\rightarrow (-u,-m)$ for the partition function $\mathcal{Z}$.

\paragraph{VEVs for Wilson loops and 1-form and 2-groups symmetries} 

In a 5D $\mathcal{N}=1$ supersymmetric gauge theory with gauge group $G$  on $\mathbb{R}^4\times S^1$, the expectation value of a Wilson loop operator in the representation $\mathbf{R}_i$ can be calculated via the localization \cite{Tong:2014cha,Nekrasov:2015wsu,Gaiotto:2015una,Kim:2016qqs,Haouzi:2020yxy} and the blowup equation \cite{Kim:2021gyj}. It takes the form
\begin{align}
    \left\langle W_{\mathbf{R}_i} \right\rangle=\mathcal{W}^{(0)}_{\mathbf{R}_i} +\sum_{k=1}^{\infty} \mathfrak{q}^k\,\mathcal{W}^{(k)}_{\mathbf{R}_i},
\end{align}
where $\mathfrak{q}$ is the instanton counting parameter that is refer to one of the mass parameters. $\mathcal{W}^{(k)}_{\mathbf{R}_i}$ is the $k$-instanton contribution, in particular, if $k=0$, it takes the expression of the character for $G$ in the representation $\mathbf{R}_i$
\begin{align}\label{eq:W0inst}
    \mathcal{W}^{(0)}_{\mathbf{R}_i}=\sum_{w \in \mathbf{R}_i}e^{w\cdot \phi},
\end{align}
with $\phi_i$ the expectation values for the scalar in the vector multiplet and $\omega$ are the weights in the representation $\mathbf{R}_i$. At the leading order, \eqref{eq:SUn_Wilson} indeed agrees with \eqref{eq:W0inst}.

It has been proposed in \cite{Huang:2022hdo}, the VEVs of the half-BPS Wilson loop operators in the 5D supersymmetric quantum field theory on $S^1$ can be calculated in its corresponding topological string B-model; for the 5D $SU(N)$ theory, at genus zero, we have \footnote{As we will see in Section~\ref{sec:examples_toric} and~\ref{sec:examples}, a more generic form of the Wilson loop is $\left\langle W_{\mathbf{R}_i} \right\rangle =u_i(1+f(z))$, where $f(z)$ is usually a finite polynomial of $z_i$. For all the pure gauge theories we will consider, we find $\left\langle W_{\mathbf{R}_i} \right\rangle=u_i+\sum_{j}P_j u_j$, where the summation is over all $u_j$'s that have the same charge as $u_i$ under the one-form symmetry. $P_j$ is a constant that only depends on mass parameters. }
\begin{align}\label{eq:defWu}
    \left\langle W_{\mathbf{R}_i} \right\rangle =u_i= \prod_{j=1}^{n-1} z_j^{-(C^{-1})_{ij}},\quad i=1,\cdots,b_4,
\end{align}
which agrees with the calculation \eqref{eq:SUn_Wilson} after performing \eqref{eq:mirrormap}. Here $\mathbf{R}_i$ is the $i^{\text{th}}$ representation\footnote{For instance, $\mathbf{R}_1=\mathbf{F}$ is the fundamental representation whose VEV is computed via $u_1$. Note that it depends on the sign notation since $u_1$ can also be related to the anti-fundamental representation $\overline{\mathbf{F}}$.}  of $G$ associated to the $i^{\text{th}}$ node of the Dynkin diagram of $SU(N)$.

Clearly, when there are no flavor branes in $M_6$, all $u_i$'s are the VEVs for genuine Wilson loops. In this case where no flavor branes is supported on any non-compact 2-cycles of $M_6$, we get:
\begin{equation}\label{eq:Extend_Gum_Gu}
    \mathcal{G}_u \cong \mathcal{E}\ (\cong \mathcal{G}_{um})\ \cong \Gamma^{(1)}.
\end{equation}
It is natural to expect that when there are flavor branes, in particular their existence breaks the 1-form symmetry, \eqref{eq:defWu} defines VEV for a ``naive'' line defect, which carries charges under $\mathcal{G}_{um}\cong\mathcal{E}$. As mentioned in Section \ref{sec:WilsonLine_M2brane}, to calculated the VEV of a Wilson loop operator from M2 branes on the relate 2-cycle $\Sigma$, we need to specify the compact representative of $\Sigma$. However, ambiguities arise from the non-uniqueness of this compact representative. Different choices of compact representatives correspond to different choices of non-compact surfaces and can be related through reparametrization of the moduli parameters $u_i$. This results in the fact that \eqref{eq:defWu} may differ by a factor that depends only on the mass parameters. As we will demonstrate in Section~\ref{sec:examples_toric} and~\ref{sec:examples}, the charge of~\eqref{eq:defWu} under $\mathcal{E}$ remains invariant under reparametrization of $u_i$, ensuring that~\eqref{eq:defWu} is a consistent observable.

\subsection{Examples: Toric cases}\label{sec:examples_toric}
In this section, we present a few toric examples about the calculation for the VEVs of Wilson loops, the one-form symmetries and two-group symmetries. Most of the examples in this section can also be found in \cite{Huang:2013yta,Haghighat:2008gw}.

\subsubsection{$SU(2)_0$}\label{sec:SU2_0}
\begin{figure}[H]
\begin{center}
\begin{tikzpicture}

\node (web) at (-5,0){
\begin{tikzpicture}[scale=0.8]
    \draw[ thick] (1,1) -- (1,-1);
    \draw[thick] (1,-1) -- (-1,-1);
    \draw[thick] (-1,-1) -- (-1,1);
    \draw[ thick] (-1,1) -- (1,1);
    \draw[thick] (1,1) -- (2.2,2.2);
    \draw[thick] (-1,1) -- (-2.2,2.2);
    \draw[thick] (1,-1) -- (2.2,-2.2);
    \draw[thick] (-1,-1) -- (-2.2,-2.2);
\end{tikzpicture}
};

\def\x{1.8}
\def\s{0}
\draw (-\x,0) -- (\x,0);
\draw (0,-\x) -- (0,\x);
\draw[thick] (-\x,0) -- (0,\x) -- (\x,0) -- (0,-\x) -- (-\x,0);

\draw (0,0) node[anchor= south west]{$u$};
\draw (-\x,0) node[anchor= east]{$m$};
\draw (-5,-2.8) node {(a)};
\draw (0,-2.8) node {(b)};
\draw (6,-2.8) node {(c)};

\node (tab) at (6,0) {%
  \begin{tabular}{c|rrr|rr}
    &  & $v_i$ &  & $Q_{1i}$ & $Q_{2i}$\\
  \hline 
  $D_u$ & $1$ & $0$ & $0$  & $-2$ &$-2$\\
  $D_1$ & $1$ & $0$ & $-1$  & $1$ &$0$\\
  $D_2$ & $1$ & $-1$ & $0$ & $0$ & $1$\\
  $D_3$ & $1$ & $0$ & $1$  & $1$ & $0$\\
  $D_4$ & $1$ & $1$ & $0$  & $0$ & $1$\\
  \end{tabular}};
\end{tikzpicture}
\end{center}
\caption{(a) Dual 5-brane web, (b) toric diagram and (c) toric data for local $\mathbb{P}^1\times \mathbb{P}^1$.}
\label{fig:toricP1P1}
\end{figure}
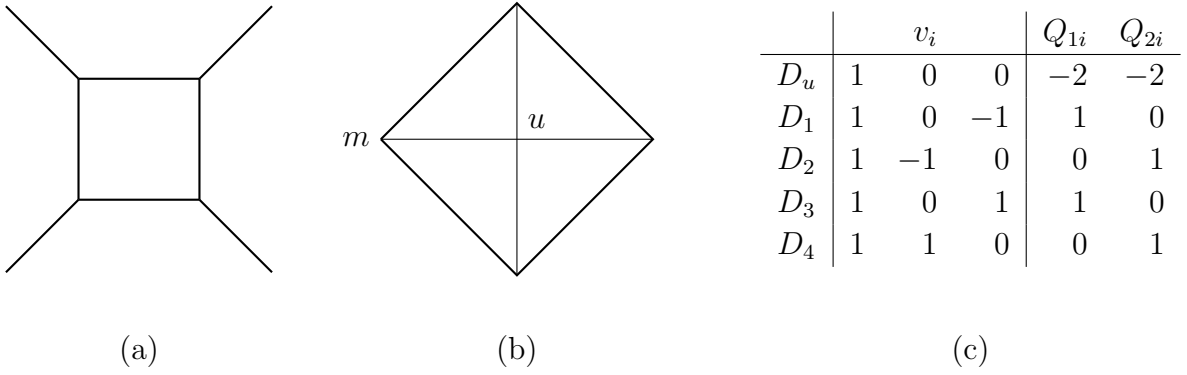

\begin{figure}[H]
\begin{center}
\begin{tikzpicture}
\node (web) at (-4.5,0){
\begin{tikzpicture}[scale=1]
    \draw[ thick] (0.5,1) -- (1.5,0);
    \draw[ thick] (0.5,1) -- (-0.5,1);
    \draw[thick] (-0.5,1) -- (-1.5,0) -- (1.5,0);
    \draw[thick] (0.5,1) -- (0.5,2);
    \draw[thick] (-0.5,1) -- (-0.5,2);
    \draw[thick] (1.5,0) -- (2.6,-0.455);
    \draw[thick] (-1.5,0) -- (-2.6,-0.455);
\end{tikzpicture}
};

\def\x{1.6}
\def\s{0}
\draw[thick] (-\x,\x) -- (\x,\x) -- (0,-\x) -- (-\x,\x);
\draw (-\x,\x) -- (0,0) -- (\x,\x);
\draw (0,\x) -- (0,-\x);

\draw (0,0) node[anchor= north west]{$u$};
\draw (-\x,\x) node[anchor= east]{$m^{\prime}$};
\draw (0,\x) node[anchor= south]{$m$};
\draw (-4.5,-2.8) node {(a)};
\draw (0,-2.8) node {(b)};
\draw (5.7,-2.8) node {(c)};

\node (tab) at (5.7,0) {%
  \begin{tabular}{c|rrr|rr}
    &  & $v_i$ &  & $Q_{1i}$ & $Q_{2i}$\\
  \hline 
  $D_u$ & $1$ & $0$ & $0$  & $-2$ &$0$\\
  $D_1$ & $1$ & $0$ & $-1$  & $1$ &$0$\\
  $D_2$ & $1$ & $-1$ & $1$ & $0$ & $1$\\
  $D_3$ & $1$ & $0$ & $1$  & $1$ & $-2$\\
  $D_4$ & $1$ & $1$ & $1$  & $0$ & $1$\\
  \end{tabular}};
  
\end{tikzpicture}
\end{center}
\caption{(a) Dual 5-brane web, (b) toric diagram and (c) toric data for local $\mathbb{F}_2$.}
\label{fig:toricF2}
\end{figure}
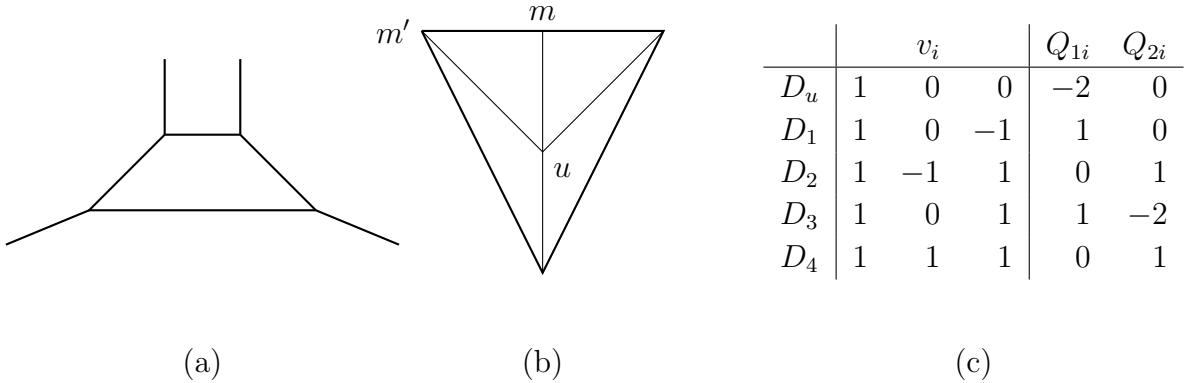
The 5D $SU(2)_0$ model is described by the local $\mathbb{F}_0$ surface, whose toric diagram is illustrated in Figure~\ref{fig:toricP1P1}. Here $D_u$ is the compact divisor and other $D_i$'s are non-compact divisors.  The invariant coordinates are given by
\begin{equation}\label{eq:F0_params}
    z_1=\frac{1}{u^2},\quad z_2=\frac{m}{u^2},
\end{equation}
where the mass parameter corresponds to the non-compact divisor $D_2$. With this choice of mass parameter, the flavor symmetry $SU(2)$ at UV is broken to $U(1)$ and we
find
$\mathcal{G}_{u}=\mathcal{G}_{um}=\mathbb{Z}_2$, which are given by
\begin{equation}
    u \mapsto -u.
\end{equation}
The VEV for the Wilson loop in the fundamental representation is $u=\frac{1}{\sqrt{z_1}}$, which can be calculated via the mirror maps. The mirror maps $t_i(z)$ for toric Calabi-Yau threefolds are periods of the B-model geometries. They are solutions to the Picard-Fuchs equations, which can be systematically constructed for toric Calabi-Yau manifolds following the methods in \cite{Candelas:1990rm,Morrison:1991cd,Batyrev:1993oya} and their generalizations \cite{Hosono:1993qy,Hosono:1994ax,Chiang:1999tz}. Typically, the mirror map can be expressed as a series expansion in the form $t_i(z)=\log z_i+\mathcal{O}(z)$. By connecting these maps to the VEVs of Wilson loops, we can expand the mirror maps in terms of instanton counting parameters, with the coefficients given in closed-form expressions. To achieve this, we first solve the mirror maps at the limit $z_2=m=0$, we have
\begin{align}
    t_1(z)|_{z_2=0}=\log\left(\frac{1-2z_1-\sqrt{1-4z_1}}{2z_1}\right), 
\end{align}
then utilize the Picard-Fuchs equations to solve the ansatz
\begin{align}
    t_1(z)=\log\left(\frac{1-2z_1-\sqrt{1-4z_1}}{2z_1}\right)+\sum_{i=1}^{\infty}f_n(z_1)z_2^n.
\end{align}
We find
\begin{align}
    t_1(z)=&-2\phi=\log\left(\frac{1-2z_1-\sqrt{1-4z_1}}{2z_1}\right)+\frac{2z_2}{(1-4z_1)^{3/2}}+\frac{3 z_2^2 \left(6 z_1+1\right)}{\left(1-4 z_1\right)^{7/2}}\nonumber\\
    &\quad\quad+\frac{20 \left(30 z_1^2+20 z_1+1\right) z_2^3}{3 \left(1-4 z_1\right)^{11/2}}+\frac{35 \left(140 z_1^3+210 z_1^2+42 z_1+1\right) z_2^4}{2 \left(1-4
   z_1\right)^{15/2}}+\mathcal{O}(z_2^5),\\
   t_2(z)=&-2\phi+\log\mathfrak{q}=\log m+t_1,
\end{align}
where $\phi$ is the Coulomb parameter and $\mathfrak{q}$ is the instanton counting parameter.
By inverting the mirror maps, we find the VEV for the Wilson loop is
\begin{align}\label{eq:u_SU2_0}
    \langle W_{\mathbf{F}}\rangle:=u=\frac{1}{\sqrt{z_1}}=&e^{\phi}+e^{-\phi}+\frac{e^{\phi}+e^{-\phi}}{(e^{\phi}-e^{-\phi})^2}\mathfrak{q}+\frac{5(e^{\phi}+e^{-\phi})}{(e^{\phi}-e^{-\phi})^6}\mathfrak{q}^2\nonumber\\
    &+\frac{(e^{\phi}+e^{-\phi})(7e^{2\phi}+7e^{-2\phi}+58)}{(e^{\phi}-e^{-\phi})^{10}}\mathfrak{q}^3\nonumber\\
    &+\frac{(e^{\phi}+e^{-\phi})(9e^{4\phi}+9e^{-4\phi}+250e^{2\phi}+250e^{-2\phi}+951)}{(e^{\phi}-e^{-\phi})^{14}}\mathfrak{q}^4+\mathcal{O}(\mathfrak{q}^5).
\end{align}
The result agrees with the localization calculation. One can indeed observe that $\langle W_{\mathbf{F}}\rangle$ defined in \eqref{eq:u_SU2_0} has charge one under the 1-form symmetry transformation $\phi\mapsto \phi+i\pi$ due to the inverse square root dependence on $z_1$ at the level of partition function.

It is well-known that the 5D $SU(2)_0$ theory can be also described by the local $\mathbb{F}_2$ geometry, whose toric diagram and toric data can be found in Figure~\ref{fig:toricF2}. If one chooses $D_3$ related to the mass parameter, the 5D theory has a manifest enhanced flavor symmetry $SU(2)$, and the invariant coordinates are given by
\begin{align}\label{eq:F2_params}
    z_1=\frac{m}{u^2},\quad z_2=\frac{1}{m^2}\,,
\end{align}
from which one can immediately read that  
$\mathcal{G}_{u}=\mathbb{Z}_2$ and $\mathcal{G}_{um}=\mathbb{Z}_4$, where the $\mathcal{G}_{um}$ action is given by
\begin{align}
    u\mapsto e^{\frac{1}{2}\pi\ri} u,\quad m\mapsto e^{\pi\ri} m\,.
\end{align}
From mirror maps
\begin{align}
    z_1=e^{-2\phi^{\prime}}\mathfrak{q}^{-\frac{1}{2}}+\cdots\,,\quad z_2=\frac{\mathfrak{q}}{(1+\mathfrak{q})^2}\,,
\end{align}
we have
\begin{align}\label{eq:u_SU2_2pi}
    u={z_1^{-\frac{1}{2}}z_2^{-\frac{1}{4}}}=e^{\phi^{\prime}}+\chi_2e^{-\phi^{\prime}}+&3e^{-3\phi^{\prime}}+5\chi_2e^{-5\phi^{\prime}}\nonumber\\
    &+(28+7\chi_3)e^{-7\phi^{\prime}}+(9\chi_4+126\chi_2)e^{-9\phi^{\prime}}+\cdots\,.
\end{align}
We find \eqref{eq:u_SU2_2pi} is the same as \eqref{eq:u_SU2_0} up to a factor:
\begin{align}
    u_{\mathbb{F}_0}(\phi,\mathfrak{q})=\mathfrak{q}^{\frac{1}{4}}\,u_{\mathbb{F}_2}(\phi^{\prime},\mathfrak{q})|_{\phi^{\prime}\rightarrow \phi-\frac{1}{2}\mathfrak{q}}\,.
\end{align}

We see that $\mathcal{G}_{um}$'s read off~(\ref{eq:F0_params}) and off~(\ref{eq:F2_params}), hence the corresponding 2-group symmetries, are not the same, though the global symmetries of the 5D SCFTs from these two compactifications must be identical in the UV. The reason is that while the 1-form symmetries remain the same across the CB~\cite{Morrison:2020ool}, the flavor symmetries can be broken in different ways on the CB, hence the middle term in the sequence~(\ref{eq:main_sequence}) can vary accordingly. Our specific choice of mass parameters for local $\mathbb{F}_0$ and for local $\mathbb{F}_2$ examplifies this point, and we are indeed calculating the global symmetries on the CB rather than in the UV. We note that there does exist a choice of the set of K\"ahler parameters that the middle element in the sequence becomes $\mathbf{Z}_4$~\cite{Closset:5D_Phases_M/IIA}, thereby matching the 2-group symmetry in the UV, though such choice is not the canoncial one on the CB.

\subsubsection{$SU(2)_{\pi}$}\label{sec:SU2_pi}
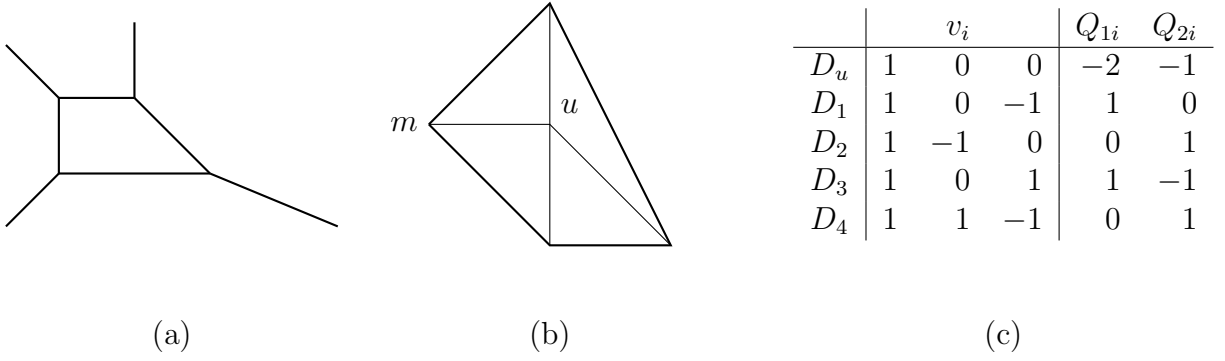
\begin{figure}[t]
\begin{center}
\begin{tikzpicture}
\node (web) at (-5,0){
\begin{tikzpicture}[scale=1]
    \draw[ thick] (0.5,1) -- (1.5,0);
    \draw[ thick] (0.5,1) -- (-0.5,1);
    \draw[thick] (-0.5,1) -- (-0.5,0) -- (1.5,0);
    \draw[thick] (0.5,1) -- (0.5,2);
    \draw[thick] (-0.5,1) -- (-1.2,1.7);
    \draw[thick] (1.5,0) -- (3.19,-0.7);
    \draw[thick] (-0.5,0) -- (-1.2,-0.7);
\end{tikzpicture}
};

\def\x{1.6}
\def\s{0}
\draw (-\x,0) -- (0,0) -- (0,-\x);
\draw (\x,-\x) -- (0,0) -- (0,\x);
\draw[thick] (-\x,0) -- (0,\x) -- (\x,-\x) -- (0,-\x) -- (-\x,0);

\draw (0,0) node[anchor= south west]{$u$};
\draw (-\x,0) node[anchor= east]{$m$};
\draw (-5,-2.8) node {(a)};
\draw (0,-2.8) node {(b)};
\draw (6,-2.8) node {(c)};

\node (tab) at (6,0) {%
  \begin{tabular}{c|rrr|rr}
    &  & $v_i$ &  & $Q_{1i}$ & $Q_{2i}$\\
  \hline 
  $D_u$ & $1$ & $0$ & $0$  & $-2$ &$-1$\\
  $D_1$ & $1$ & $0$ & $-1$  & $1$ &$0$\\
  $D_2$ & $1$ & $-1$ & $0$ & $0$ & $1$\\
  $D_3$ & $1$ & $0$ & $1$  & $1$ & $-1$\\
  $D_4$ & $1$ & $1$ & $-1$  & $0$ & $1$\\
  \end{tabular}};
  
\end{tikzpicture}
\end{center}
\caption{(a) Dual 5-brane web, (b) toric diagram and (c) toric data for local $\mathbb{F}_1$.}
\label{fig:toricF1}
\end{figure}
The 5D $SU(2)_{\pi}$ model is described by the local $\mathbb{F}_1$, whose toric diagram is illustrated in Figure~\ref{fig:toricF1}. Here $D_u$ is the compact divisor and other $D_i$'s are non-compact divisors.  The invariant coordinates are given by
\begin{equation}
    z_1=\frac{1}{u^2},\quad z_2=\frac{m}{u}
\end{equation}
from which we can read that there is no 1-form symmetry and 
$\mathcal{G}_{um}=\mathbb{Z}_2$ acts as
\begin{equation}
    u \mapsto -u,\quad m \mapsto -m.
\end{equation}
To compute the instanton expansion for the Wilson loop, we first compute the mirror maps
\begin{align}
    t_1=&-2\phi=\log\left(\frac{1-2z_1-\sqrt{1-4z_1}}{2z_1}\right)-\frac{4 z_1z_2}{\left(1-4 z_1\right)^{3/2}}+\frac{30 z_1^2 z_2^2}{\left(1-4 z_1\right)^{7/2}}\nonumber\\
    &\quad\quad\quad\quad\quad\quad\quad\quad\quad\quad-\frac{560  \left(2+z_1\right) z_1^3z_2^3}{3 \left(1-4 z_1\right)^{11/2}}+\frac{1155  \left(5+6 z_1\right) z_1^4z_2^4}{\left(1-4 z_1\right)^{15/2}}+\mathcal{O}(z_2^5),\\
    t_2=&-\phi+\log\frak{q}=\log m+\frac{1}{2}t_1,
\end{align}
via the Picard-Fuchs equations. We find the VEV of the Wilson loop in the fundamenal representation takes the expression
\begin{align}\label{eq:W_SU2}
    \langle W_{\mathbf{F}}\rangle:=u&=\frac{1}{\sqrt{z_1}}=e^{\phi}+e^{-\phi}-\frac{2}{(e^{\phi}-e^{-\phi})^2}\mathfrak{q}+\frac{5(e^{\phi}+e^{-\phi})}{(e^{\phi}-e^{-\phi})^6}\mathfrak{q}^2\nonumber\\
    &\quad-\frac{80+32e^{2\phi}+32e^{-2\phi}}{(e^{\phi}-e^{-\phi})^{10}}\mathfrak{q}^3+\frac{13(e^{\phi}+e^{-\phi})(22e^{2\phi}+22e^{-2\phi}+69)}{(e^{\phi}-e^{-\phi})^{14}}\mathfrak{q}^4+\mathcal{O}(\mathfrak{q}^5).
\end{align}
At the level of the partition function, the 1-form symmetry action $\phi\mapsto \phi+i\pi$ is broken, and the Wilson loop \eqref{eq:W_SU2} does not obtain an overall phase under the action. However, if one performs the action $\phi\mapsto \phi+i\pi,\,\frak{q}\mapsto -\frak{q} \in  \mathcal{G}_{um}\cong \mathcal{E}=\mathbb{Z}_2$, the Wilson loop $\langle W_{\mathbf{F}}\rangle$ calculated in \eqref{eq:W_SU2} has an overall phase $e^{i\pi}$. 
\subsubsection{$SU(3)_{0}$}
\begin{figure}[t]
\begin{center}
\begin{tikzpicture}

\node (web) at (-9,-1){
\begin{tikzpicture}[scale=1]
    \draw[ thick] (0.5,1) -- (1.5,0);
    \draw[ thick] (0.5,1) -- (-0.5,1);
    \draw[thick] (-0.5,1) -- (-0.5,0) -- (1.5,0);
    \draw[thick] (0.5,1) -- (0.5,2);
    \draw[thick] (-0.5,1) -- (-1.5,2);
    \draw[thick] (0.5,2) -- (-1.5,2);
    \draw[thick] (1.5,0) -- (3.19,-0.7);
    \draw[thick] (-0.5,0) -- (-1.2,-0.7);
    \draw[thick] (0.5,2) -- (1.2,2.7);
    \draw[thick] (-1.5,2) -- (-3.19,2.7);
\end{tikzpicture}
};
\def\x{1.6}
\def\s{0}
\draw (0,-\x) -- (0,\x);
\draw (-\x,-\x) -- (0,0) -- (\x,0);
\draw[thick] (0,-2*\x) -- (\x,0) -- (0,\x) -- (-\x,-\x) -- (0,-2*\x);
\draw (0,-2*\x) -- (0,\x);
\draw (-\x,-\x) -- (0,-\x) -- (\x,0);
\draw (0,0) node[anchor= south west]{$u_1$};
\draw (0,-\x) node[anchor= south east]{$u_2$};
\draw (-\x,-\x) node[anchor= east]{$m$};
\draw (-8.6,-4.2) node {(a)};
\draw (0,-4.2) node {(b)};
\draw (-4.4,-10) node {(c)};

\node (tab) at (-4.4,-7.2) {%
  \begin{tabular}{c|rrr|rrr}
    &  & $v_i$ &  & $Q_{1i}$ & $Q_{2i}$ & $Q_{3i}$\\
  \hline 
  $D_1$ & $1$ & $0$ & $1$  & $1$ &$0$ & $0$\\
  $D_{u_1}$ & $1$ & $0$ & $0$ & $-2$ & $1$ & $-1$\\
  $D_{u_2}$ & $1$ & $0$ & $-1$  & $1$ & $-2$ & $-1$\\
  $D_2$ & $1$ & $0$ & $-2$  & $0$ &$1$ &  $0$\\
  $D_3$ & $1$ & $-1$ & $-1$  & $0$ & $0$ & $1$\\
  $D_4$ & $1$ & $1$ & $0$  & $0$ & $0$ & $1$\\
  \end{tabular}};
\end{tikzpicture}
\end{center}
\caption{(a) Dual 5-brane web, (b) toric diagram and (c) toric data for the Sasaki–Einstein manifold $Y^{3,0}$.}
\label{fig:toricSU3}
\end{figure}
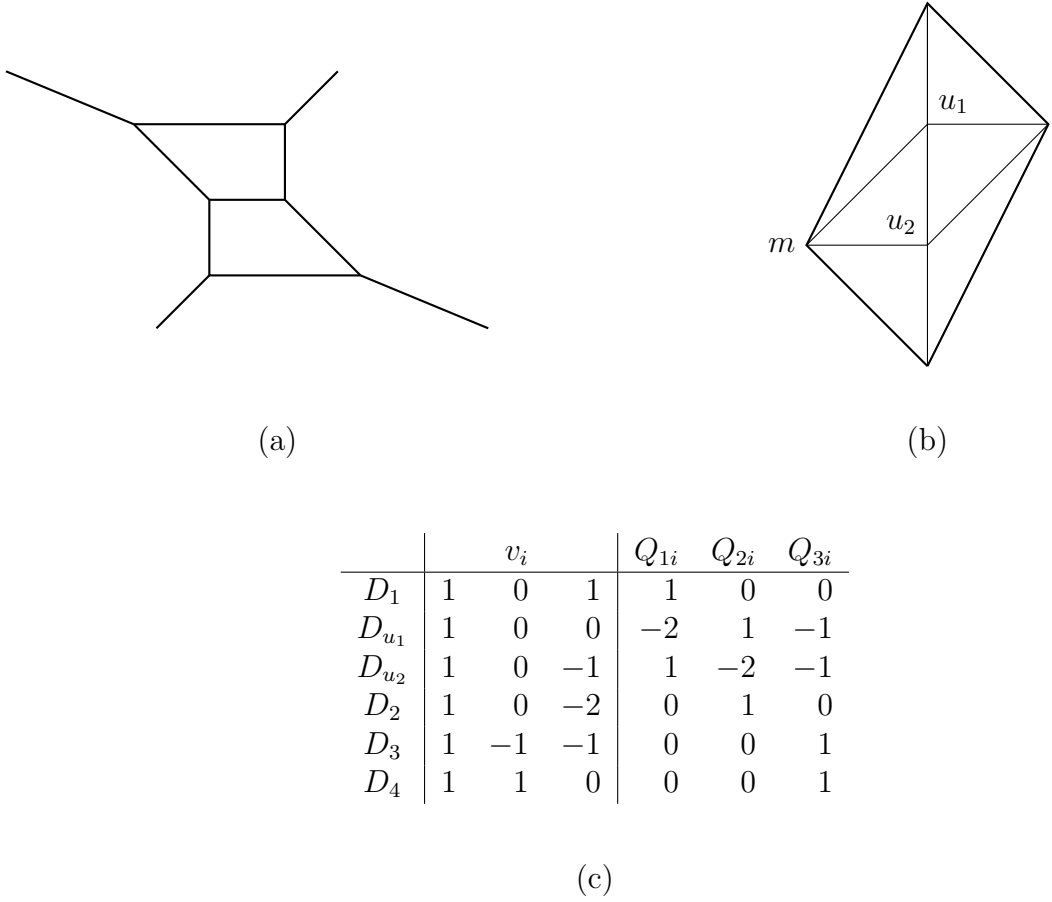
The 5D $SU(3)_{0}$ theory is described by the local geometry, the
Sasaki–Einstein manifold $Y^{3,0}$, illustrated in Figure~\ref{fig:toricSU3}. Here $D_{u_1}$ and $D_{u_2}$ are compact divisors and other $D_i$'s are non-compact divisors.  The invariant coordinates are given by
\begin{equation}
    z_1=\frac{u_1}{u_2^2},\quad z_2=\frac{u_2}{u_1^2},\quad z_3=\frac{m }{u_1 u_2}.
\end{equation}
One can read that there is a
$\mathcal{G}_u=\mathcal{G}_{um}=\mathbb{Z}_3$ symmetry
\begin{equation}
    u_1 \mapsto e^{\frac{4}{3}\pi \ri  }u_1,\quad u_2 \mapsto e^{\frac{2}{3}\pi \ri  }u_2,
\end{equation}
which leaves the coordinates $z_i$ invariant. 

To compute the mirror maps, we first propose the ansatz
\begin{align}\label{eq:ansatz_SU3}
    t_i=\log z_i+\mathcal{O}(z)=f_{0,i}(z_1,z_2)+\sum_{n=0}^{\infty}f_{n,i}(z_1,z_2)z_3^n,\quad i=1,2.
\end{align}
By utilizing the Picard-Fuchs equations, we find $f_{0,1},f_{0,2}$ can be solved from the equations
\begin{align}\label{eq:ztof_SU3}
    z_1=\frac{e^{f_{0,1}}(1+e^{f_{0,2}}+e^{f_{0,1}+f_{0,2}})}{(1+e^{f_{0,1}}+e^{f_{0,1}+f_{0,2}})^2},\quad z_2=\frac{e^{f_{0,2}}(1+e^{f_{0,1}}+e^{f_{0,1}+f_{0,2}})}{(1+e^{f_{0,2}}+e^{f_{0,1}+f_{0,2}})^2}.
\end{align}
By taking $z_j$ derivatives on both sides for the equations in  \eqref{eq:ztof_SU3}, we obtain the derivative rules for $\partial_{z_j}f_{0,i},i,j=1,2$, as functions of $f_{0,1},f_{0,2}$.
All other $f_{n,1},f_{n,2}$ can be solved recursively from the derivative rules and the third Picard-Fuchs equation
\begin{align}
    \mathcal{L}_3=\theta _3^2+z_3 \left(\theta _1-2 \theta _2-\theta _3\right) \left(2 \theta _1-\theta
   _2+\theta _3\right),
\end{align}
where $\theta_i=z_i\partial_{z_i}$.

After fixing the ansatz \eqref{eq:ansatz_SU3}, the Wilson loops are given by
\begin{align}
    \langle W_{\mathbf{F}}\rangle:=&\,\,u_1=e^{\phi _1}+e^{-\phi _2}+e^{-\phi _1+\phi _2}\nonumber\\
    &+\frac{2 e^{2 \left(\phi
   _1+\phi _2\right)} \left(e^{4 \phi _1}+e^{2 \phi _2}-e^{2 \phi _1+\phi _2}-e^{3 \phi
   _1+2 \phi _2}-e^{\phi _1+3 \phi _2}+e^{2 \phi _1+4 \phi _2}\right) \frak{q}}{\left(e^{2 \phi
   _1}-e^{\phi _2}\right)^2 \left(e^{\phi _1}-e^{2 \phi _2}\right)^2
   \left(-1+e^{\phi _1+\phi _2}\right)^2}+\mathcal{O}(\frak{q}^2),\\
    \langle W_{\overline{\mathbf{F}}}\rangle:=&\,\,u_2=\left(e^{-\phi _1}+e^{\phi _1-\phi _2}+e^{\phi _2}\right)\nonumber\\
    &+\frac{2 e^{2 \left(\phi
   _1+\phi _2\right)} \left(e^{2 \phi _1}+e^{4 \phi _2}-e^{3 \phi _1+\phi _2}-e^{\phi
   _1+2 \phi _2}+e^{4 \phi _1+2 \phi _2}-e^{2 \phi _1+3 \phi _2}\right) \frak{q}}{\left(e^{2
   \phi _1}-e^{\phi _2}\right)^2 \left(e^{\phi _1}-e^{2 \phi _2}\right)^2
   \left(-1+e^{\phi _1+\phi _2}\right)^2}+\mathcal{O}(\frak{q}^2),
\end{align}
which are consistent with the results from localization calculations.

\subsubsection{$SU(2)_{0}\times SU(2)_0$}

\begin{figure}[t]
\begin{center}
\begin{tikzpicture}
\node (web) at (-9,0){
\begin{tikzpicture}[scale=1]
    \draw[ thick] (0.5,1) -- (1,0.5) ;
    \draw[ thick] (1,0.5) -- (1,-0.2);
    \draw[ thick] (0.5,1) -- (-0.5,1);
    \draw[thick] (-0.5,1) -- (-1.7,-0.2) -- (1,-0.2);
    \draw[thick] (0.5,1) -- (0.5,2);
    \draw[thick] (-0.5,1) -- (-0.5,2);
    \draw[thick] (1,-0.2) -- (1.8,-1);
    \draw[thick] (3.4,-1) -- (1.8,-1);
    \draw[thick] (3.4,-1) -- (1.9,0.5);
    \draw[thick] (1,0.5) -- (1.9,0.5);
    \draw[thick] (-1.7,-0.2) -- (-2.8,-0.655);
    \draw[thick] (3.4,-1) -- (4.5,-1.455);
    \draw[thick] (1.8,-1) -- (1.8,-1.7);
    \draw[thick] (1.9,0.5) -- (1.9,2);
\end{tikzpicture}
};
\def\x{1.6}
\def\s{0}
\draw (-\x,-\x) -- (0,0) -- (0,\x);
\draw (-\x,0) -- (0,0) -- (\x,\x);
\draw[thick] (-2*\x,\x) -- (-\x,\x) -- (0,\x) -- (\x,\x) -- (0,-\x)-- (-\x,-\x) -- (-2*\x,\x);
\draw (-2*\x,\x) -- (-\x,0);
\draw (-\x,-\x) -- (-\x,0) -- (0,\x);
\draw (-\x,0) -- (-\x,\x);
\draw (0,0) -- (0,-\x);
\draw (0,0) node[anchor= north west]{$u_2$};
\draw (-\x,0) node[anchor= north east]{$u_1$};
\draw (-1*\x,\x) node[anchor= south]{$m_1$};
\draw (0*\x,\x) node[anchor= south]{$m_2$};
\draw (0*\x,-\x) node[anchor= north]{$m_3$};

\draw (-8.5,-2.7) node {(a)};
\draw (-1,-2.7) node {(b)};
\draw (-5,-9.7) node {(c)};

\node (tab) at (-5.4,-6.4) {%
  \begin{tabular}{c|rrr|rrrrr}
    &  & $v_i$ &  & $Q_{1i}$ & $Q_{2i}$ & $Q_{3i}$ & $Q_{4i}$ & $Q_{5i}$\\
  \hline 
  $D_{u_1}$ & $1$ & $-1$ & $0$  & $0$ &$-1$ & $-1$ & $1$ & $1$\\
  $D_{u_2}$ & $1$ & $0$ & $0$  & $0$ &$1$ & $-1$ & $-1$ & $-1$\\
  $D_1$ & $1$ & $0$ & $1$ & $1$ & $-1$ & $1$ & $-1$ & $0$\\
  $D_2$ & $1$ & $0$ & $-1$  & $0$ & $0$ & $0$ & $0$ & $1$\\
  $D_3$ & $1$ & $-1$ & $1$  & $-2$ & $1$ & $0$ &$0$ & $0$\\
  $D_4$ & $1$ & $-1$ & $-1$  & $0$ & $0$ & $1$ &$0$ & $-1$\\
  $D_5$ & $1$ & $-2$ & $1$  & $1$ & $0$ & $0$ & $0$ & $0$\\
  $D_6$ & $1$ & $1$ & $1$  & $0$ & $0$ & $0$ & $1$ & $0$\\
  \end{tabular}};

\end{tikzpicture}
\end{center}
\caption{(a) Dual 5-brane web, (b) toric diagram and (c) toric data for 5D $SU(2)_0\times SU(2)_0$ theory.}
\label{fig:toricSU2SU2}
\end{figure}
The 5D $SU(2)_{0}\times SU(2)_0$ theory is described by the local geometry illustrated in Figure~\ref{fig:toricSU2SU2}. Here $D_{u_1}$ and $D_{u_2}$ are compact divisors and other $D_i$'s are non-compact divisors. 
The brane diagram has three external parallel branes, indicating that
the theory has a $\mathcal{F}=SU(3)$ global symmetry. The invariant coordinates that manifest the $SU(3)$ global symmetry are
\begin{equation}
    z_1=\frac{m_2}{m_1^2},\quad  z_2=\frac{m_1u_2}{m_2u_1},\quad z_3=\frac{m_2}{u_1u_2 },\quad z_4=\frac{ u_1}{m_2u_2},\quad z_5=\frac{m_3u_1}{u_2}
\end{equation}
We find  the symmetry group $\mathcal{G}_u=\mathbb{Z}_2$ acts as
\begin{align}
    u_1\mapsto -u_1,\quad u_2\mapsto -u_2,
\end{align}
and the
$\mathcal{G}_{um}=\mathbb{Z}_6$ symmetry group acts as
\begin{align}
    u_1\mapsto e^{\frac{\pi i}{3}}u_1,\quad u_2\mapsto e^{{\pi i}}u_2,\quad m_1\mapsto e^{\frac{2\pi i}{3}}m_1,\quad m_2\mapsto e^{\frac{4\pi i}{3}}m_2,\quad m_3\mapsto e^{\frac{2\pi i}{3}}m_3.
\end{align}
From $\mathcal{G}_{um}$ and $\mathcal{G}_{u}$, we obtain the short exact sequence
\begin{align}
    0\rightarrow\mathcal{F}\rightarrow\mathbb{Z}_6\rightarrow \mathbb{Z}_2\rightarrow 0,
\end{align}
indicating the global form of the flavor symmetry is
\begin{align}
    PSU(3)\cong SU(3)/\mathbb{Z}_3
\end{align}

\subsubsection{$SU(2)+\mathbf{F}$}\label{sec:SU2+1F}

\begin{figure}[t]
\begin{center}
\begin{tikzpicture}
\node (web) at (-3.9,0){
\begin{tikzpicture}[scale=1]
    \draw[ thick] (0.5,1) -- (0.9,0.6) -- (0.9,0);
    \draw[ thick] (0.5,1) -- (-0.5,1);
    \draw[thick] (-0.5,1) -- (-1.5,0) -- (0.9,0);
    \draw[thick] (0.5,1) -- (0.5,2);
    \draw[thick] (-0.5,1) -- (-0.5,2);
    \draw[thick] (0.9,0) -- (1.4,-0.5);
    \draw[thick] (0.9,0.6) -- (1.47,0.6);
    \draw[thick] (-1.5,0) -- (-2.6,-0.455);
\end{tikzpicture}
};

\def\x{1.6}
\def\s{0}
\draw[thick] (-\x,\x) -- (\x,\x) -- (\x,0) -- (0,-\x) -- (-\x,\x);
\draw (-\x,\x) -- (0,0) -- (\x,\x);
\draw (0,0) -- (\x,0);
\draw (0,\x) -- (0,-\x);

\draw (0,0) node[anchor= north west]{$u$};
\draw (0,\x) node[anchor= south]{$m_1$};
\draw (\x,0) node[anchor= west]{$m_2$};
\draw (-3.9,-2.8) node {(a)};
\draw (0,-2.8) node {(b)};
\draw (6,-2.8) node {(c)};

\node (tab) at (6,0) {%
  \begin{tabular}{c|rrr|rrr}
    &  & $v_i$ &  & $Q_{1i}$ & $Q_{2i}$ & $Q_{3i}$\\
  \hline 
  $D_u$ & $1$ & $0$ & $0$  & $-1$ &$0$ &$-1$\\
  $D_1$ & $1$ & $0$ & $-1$  & $0$ &$0$ &$1$\\
  $D_2$ & $1$ & $-1$ & $1$ & $0$ & $1$ &$0$\\
  $D_3$ & $1$ & $0$ & $1$  & $1$ & $-2$ &$0$\\
  $D_4$ & $1$ & $1$ & $1$  & $-1$ & $1$ &$1$\\
  $D_4$ & $1$ & $1$ & $0$  & $1$ & $0$ &$-1$\\
  \end{tabular}};
  
\end{tikzpicture}
\end{center}
\caption{(a) Dual 5-brane web, (b) toric diagram and (c) toric data for $dP_2$.}
\label{fig:toricdP2_2}
\end{figure}
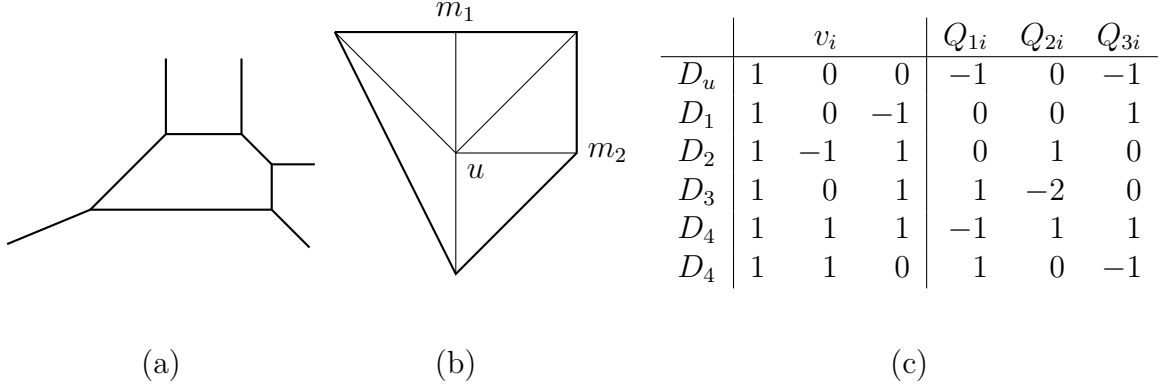
The 5D $SU(2)+\mathbf{F}$ theory corresponding to the geometry in Figure \ref{fig:toricdP2_2} has an enhanced flavor symmetry $SU(2)\times U(1)$, with the flavor fugacities $m_1$ and $m_2$. The invariant coordinates are expressed as
\begin{align}
    z_1=\frac{m_1m_2}{u},\quad z_2=\frac{1}{m_1^2},\quad z_3=\frac{1}{m_2 u},
\end{align}
from which we find $\mathcal{G}_{u}$ is trivial and $\mathcal{G}_{um}=\mathbb{Z}_4$ generated by
\begin{align}\label{eq:act_SU2F}
    u\mapsto e^{\frac{\pi i}{2}} u,\quad m_2\mapsto e^{-\frac{\pi i}{2}} m_2,\quad m_1\mapsto e^{-\pi i} m_1.
\end{align}
However, if we restrict the non-Abalien part of the flavor symmetry as has been done in \cite{Apruzzi:2021vcu} by requiring that the mass parameter $m_1$ has no phase under the transformation, we obtain trivial $\mathcal{G}_{um}$ hence $\mathcal{E}=\emptyset$.

As an additional verification of the $\mathcal{G}_{um}=\mathbb{Z}_4$ symmetry, we observe that the superconformal index for the $SU(2)+\mathbf{F}$ theory, calculated in \cite[(4.11)]{Kim:2012gu}, is indeed invariant under the action \eqref{eq:act_SU2F}.

\subsection{Examples: Non-Toric cases}\label{sec:examples}
In this section, we present two non-toric examples that can be constructed as hypersurfaces in toric varieties. We focus on  calculations related to the VEVs of Wilson loops, the one-form symmetries and the two-group symmetries. 
\subsubsection{$SO(5)$}\label{sec:B2}
\begin{equation} \label{B2polytope1}
 \begin{tabular}{c|rrrr|rrrrr|} 
    &&\multicolumn{2}{c}{$\,v_i$}&    & $Q_{1i}$ & $Q_{2i}$ & $Q_{3i}$ &  \\ \hline  
    $D_{1}$& $0$ & $0$ & $0$ & $0$ & $-2$ & $0$ & $0$ & \\
$D_{2}$& $-1$ & $0$ & $0$ & $0$ & $0$ & $1$ & $0$ & \\
$D_{3}$& $1$ & $2$ & $1$ & $0$ & $0$ & $0$ & $1$ & \\
$D_{4}$& $0$ & $-1$ & $0$ & $0$ & $1$ & $0$ & $0$ & \\
$D_{5}$& $2$ & $3$ & $0$ & $-2$ & $1$ & $0$ & $-1$ & \\
$D_{u_1}$& $1$ & $2$ & $0$ & $-2$ & $-2$ & $1$ & $-1$ & \\
$D_{u_2}$& $0$ & $1$ & $0$ & $-1$ & $2$ & $-2$ & $0$ & \\
$D_{6}$& $2$ & $3$ & $-1$ & $-4$ & $0$ & $0$ & $1$ & \\
\end{tabular} \  
\end{equation} 
The geometry for 5D $SO(5)$ theory is not a toric CY3. However, it can be constructed as a hypersurface in a non-compact toric variety. This is achieved by taking the non-compact limit of a compact Calabi-Yau hypersurface. For example, by removing the rays $r_z,r_u,r_v,r_g,r_d$ in \eqref{eq:raysSO9}, we obtain the polytope in \eqref{B2polytope1} which describes the geometry for the 5D $SO(5)$ theory. The Mori cone matrix in \eqref{B2polytope1} defines the invariant coordinates:
\begin{equation}
    z_1=\frac{u_2^2}{u_1^2},\quad  z_2=\frac{u_1}{u_2^2},\quad z_3=\frac{m}{u_1},
\end{equation}
where $m$ corresponds to the non-compact surface $D_6$ or $D_3$. From these coordinates, we observe that $\mathcal{G}_u=\mathcal{G}_{um}=\mathbb{Z}_2$, 
where the action is given by
\begin{align}
    u_2\mapsto - u_2.
\end{align}
This indicates the one-form symmetry is $\Gamma^{(1)}=\mathbb{Z}_2$.

By calculating the mirror maps from the Picard-Fuchs equations, we find the VEVs for the Wilson loops are
\begin{align}
    \langle W_{\mathbf{F}}\rangle:=&\,\,u_1+1=1+e^{-\phi _1}+e^{\phi _1}+e^{\phi _1-2 \phi _2}+e^{-\phi _1+2 \phi _2}\nonumber\\
    &\!\!\!\!\!\!\!\!\!\!\!\!\!\!\!+\frac{2 e^{2
   \phi _2} \left(e^{2 \phi _1}+e^{3 \phi _1}+e^{4 \phi _1}+e^{4 \phi _2}+e^{\phi _1+2
   \phi _2}+e^{3 \phi _1+2 \phi _2}+e^{\phi _1+4 \phi _2}+e^{2 \phi _1+4 \phi _2}\right)
   \frak{q}}{\left(e^{2 \phi _1}-e^{2 \phi _2}\right)^2 \left(-1+e^{2 \phi _2}\right)^2}+\mathcal{O}(\frak{q}^2)\,,\\
   \langle W_{\mathbf{4}}\rangle:=&\,\,u_2=e^{\phi _1-\phi _2}+e^{-\phi _2}+e^{\phi _2}+e^{-\phi _1+\phi _2}+\frac{2 e^{\phi _1+3
   \phi _2} \left(1+e^{\phi _1}\right) \left(e^{\phi _1}+e^{2 \phi _2}\right)
   \frak{q}}{\left(e^{2 \phi _1}-e^{2 \phi _2}\right)^2 \left(-1+e^{2 \phi _2}\right)^2}+\mathcal{O}(\frak{q}^2)\,,
\end{align}
they have charge zero and charge one under the one-form symmetry $\mathbb{Z}_2$.

\subsubsection{$G_2$ vs. $SU(3)_7$}\label{sec:G2}

\begin{equation} \label{G2polytope1}
 \begin{tabular}{c|rrrr|rrrrr|} 
    &&\multicolumn{2}{c}{$\,v_i$}&    & $Q_{1i}$ & $Q_{2i}$ & $Q_{3i}$ &  \\ \hline  
    $D_{1}$& $0$ & $0$ & $0$ & $0$ & $-3$ & $0$ & $-1$ & \\
$D_{2}$& $-1$ & $0$ & $0$ & $0$ & $1$ & $0$ & $0$ & \\
$D_{3}$& $0$ & $-1$ & $0$ & $0$ & $0$ & $1$ & $0$ & \\
$D_{4}$& $2$ & $3$ & $1$ & $0$ & $0$ & $0$ & $1$ & \\
$D_{5}$& $2$ & $3$ & $0$ & $-1$ & $1$ & $0$ & $0$ & \\
$D_{u_1}$& $2$ & $3$ & $0$ & $-2$ & $-2$ & $1$ & $-2$ & \\
$D_{u_2}$& $1$ & $1$ & $0$ & $-1$ & $3$ & $-2$ & $1$ & \\
$D_{6}$& $1$ & $2$ & $-1$ & $-3$ & $0$ & $0$ & $1$ & \\
\end{tabular} \  
\end{equation} 
The geometry of the 5D pure $G_2$ gauge theory can be obtained from a non-compact limit of the compact elliptic-fibered CY3 described in \ref{sec:geom_G2}, by removing the rays $r_z,r_p,r_d$. This local geometry is a Calabi-Yau hypersurface whose toric data is in equation \eqref{G2polytope1}. The invariant coordinates are
\begin{equation}\label{eq:G2zu1}
    z_1=\frac{u_2^3}{u_1^2},\quad  z_2=\frac{u_1}{u_2^2},\quad z_3=\frac{mu_2}{u_1^2},
\end{equation}
where $m$ is related to the non-compact surface $D_6$ or $D_4$. Under this parameterization, the parameter $m$ is the instanton counting parameter for the 5D pure $G_2$ theory. One can observe the groups $\mathcal{G}_{u}$ and $\mathcal{G}_{um}$ are trivial.

By computing the mirror maps, we find the VEVs for Wilson loops in the fundamental representation $\mathbf{7}$ and the adjoint representation $\mathbf{14}$ are
\begin{align}\label{eq:wilsonG2}
  \langle W_{\mathbf{7}}^{G_2}\rangle:=&\,\,u_1+7=\chi_{\mathbf{7}}-\frac{\chi_{\mathbf{14}}+3\chi_{\mathbf{7}}+1}{(1-e^{\pm \phi_1})(1-e^{\pm (2\phi_1-3\phi_2)})(1-e^{\pm (-\phi_1+3\phi_2)})}\mathfrak{q}+\mathcal{O}(\frak{q}^2),\\
   \langle W_{\mathbf{14}}^{G_2}\rangle:=&\,\,u_2+4u_1+14=\chi_{\mathbf{14}}-\frac{\chi_{\mathbf{64}}+2\chi_{\mathbf{14}}+6\chi_{\mathbf{7}}+10}{(1-e^{\pm \phi_1})(1-e^{\pm (2\phi_1-3\phi_2)})(1-e^{\pm (-\phi_1+3\phi_2)})}\mathfrak{q}+\mathcal{O}(\frak{q}^2),
\end{align}
where $\chi_{\mathrm{dim}\,\mathbf{R}}$ is the character of $G_2$ in the representation $\mathbf{R}$. They are related to the Coulomb parameters via
\begin{align}
    \chi_{\mathbf{7}}&=1+e^{\phi _1-2 \phi _2}+e^{\phi _1-\phi _2}+e^{-\phi _2}+e^{\phi _2}+e^{-\phi _1+\phi
   _2}+e^{-\phi _1+2 \phi _2},\\
   \chi_{\mathbf{14}}&=1+e^{-\phi _1}+e^{\phi _1}+e^{\phi _1-3 \phi _2}+e^{2 \phi _1-3 \phi _2}+e^{-2 \phi _1+3
   \phi _2}+e^{-\phi _1+3 \phi _2}+\chi_{\mathbf{7}}
\end{align}
and $\chi_{\mathbf{64}}=\chi_{\mathbf{7}}\chi_{\mathbf{14}}-\chi_{\mathbf{7}}^2+\chi_{\mathbf{14}}+1$\,.

\vspace{1cm}

It is known that the 5D $G_2$ theory is dual to the 5D $SU(3)_7$ theory with Chern-Simons level 7 \cite{Bhardwaj:2019ngx,Bhardwaj:2020gyu}. The two theories have a common UV completion and they are described by the same geometry but different parameterization. The parameterization for the 5D $SU(3)_7$ theory is 
\begin{equation}
    z_1=\frac{\tilde{m}\tilde{u}_2^3}{\tilde{u}_1^2},\quad  z_2=\frac{\tilde{u}_1}{\tilde{u}_2^2},\quad z_3=\frac{\tilde{u}_2}{\tilde{u}_1^2},
\end{equation}
where $\tilde{m}=\tilde{\mathfrak{q}}$ is the instanton counting parameter for the $SU(3)$ theory. The $SU(3)$ gauge group has a $\mathbb{Z}_3$ center, which is broken for $SU(3)_7$ theory; however, a $\mathbb{Z}_3$ symmetry can be restored if we consider the two group action $\tilde{\mathcal{G}}_{um}=\mathbb{Z}_3$:
\begin{align}\label{eq:G2zu2}
    \tilde{u}_2\mapsto e^{\frac{2\pi i}{3}},\quad \tilde{u}_1\mapsto e^{\frac{4\pi i}{3}},\quad \tilde{m}\mapsto  e^{\frac{2\pi i}{3}} \tilde{m}
\end{align}
which means instanton counting parameter $\tilde{\mathfrak{q}}$ has charge $1$ under $\tilde{\mathcal{G}}_{um}$. By computing the mirror maps, we find the VEVs for the Wilson loops in the fundamental and anti-fundamental representations are
\begin{align}
  \langle W_{\mathbf{3}}^{SU(3)_7}\rangle:=&\,\,\tilde{u}_2=\chi_{(10)}-\frac{-\chi_{(41)}+9\chi_{(22)}+38\chi_{(11)}-19\chi_{30}-22\chi_{(03)}-102}{(1-e^{\pm (\phi_1+\phi_2)})(1-e^{\pm (-\phi_1+2\phi_2)})(1-e^{\pm (2\phi_1-\phi_2)})}\tilde{\mathfrak{q}}+\mathcal{O}(\tilde{\mathfrak{q}}{}^2),\\
   \langle W_{\overline{\mathbf{3}}}^{SU(3)_7}\rangle:=&\,\,\tilde{u}_1=\chi_{(01)}\nonumber\\
   &\!\!\!\!\!\!\!\!\!\!\!\!\!\!\!\!\!\!\!\!\!\!\!\!\!\!\!-\frac{-\chi_{(51)}+6\chi_{(32)}-9\chi_{(13)}+2\chi_{(21)}-10\chi_{(40)}+10\chi_{(21)}+14\chi_{(02)}-34\chi_{(10)}}{(1-e^{\pm \phi_1})(1-e^{\pm (2\phi_1-3\phi_2)})(1-e^{\pm (-\phi_1+3\phi_2)})}\tilde{\mathfrak{q}}+\mathcal{O}(\tilde{\mathfrak{q}}{}^2),
\end{align}
they have charges $1$ and $2$ under the action $\tilde{\mathcal{G}}_{um}$.

We expect that the Wilson loops in two different descriptions are related, and they should have the same charge under the action $\mathcal{G}_{um}$ and $\tilde{\mathcal{G}}_{um}$. However, naively, one may see their charges under the group action in $\mathcal{G}_{um}$ are different. By requiring that the invariant coordinates are the same in \eqref{eq:G2zu1} and \eqref{eq:G2zu2}, we find 
\begin{align}\label{eq:G2_utout}
    u_1=\frac{\tilde{u}_1}{\tilde{m}^2},\quad u_2=\frac{\tilde{u}_2}{\tilde{m}},\quad m=\frac{1}{\tilde{m}^3}.
\end{align}
From \eqref{eq:G2_utout}, we find $u_1,u_2$ and $m$ has charge zero under $\tilde{\mathcal{G}}_{um}$, which indicates that the Wilson loops calculated in \eqref{eq:wilsonG2} are indeed charge 0 under the 2-group action $\mathcal{G}_{um}$.

\section{Gauging, decoupling and a conjecture relating 2-group symmetry to Coulomb branch geometry}

Having obtained the key short exact sequence~(\ref{eq:Extend_Gum_Gu}), we discuss some of its physical consequences in Section~\ref{sec:phys_consequences}. We will conjecture a relation between the torsion subgroup of the Mordell-Weil group of the special geometry of the SCFT and its 2-group symmetry in Section~\ref{sec:conjecture}.

\subsection{Gauging, decoupling and a check via rank-1 theories}\label{sec:phys_consequences}

\paragraph{Gauging and decoupling}

An immediate consequence of the above discussion of $\mathcal{G}_{um}$ and $\mathcal{G}_u$ is that $\mathcal{G}_u$ can change when some of the mass parameters are converted to Coulomb parameters. Practically this means that turning certain $m$'s into $u$'s in $z = z(u,m)$ and geometrically this means that certain non-compact divisors are compactified, causing the corresponding non-dynamical $U(1)$'s to be gauged. We denote the new set of parameters by $u'$, which by definition satisfies $|\mathcal{G}_{u'}| \geq |\mathcal{G}_{u}|$.

\begin{figure}[t]
\begin{center}
\begin{tikzpicture}
\node (web) at (-3.9,0){
\begin{tikzpicture}[scale=1.5]
    \draw[ thick] (0,0) -- (0,-2);
    \draw[ thick] (0,0) -- (0,2);
    \draw[ thick] (-1,0) -- (1,0);
    \draw[ thick] (-1,0) -- (0,-1) -- (1,0);
    \draw[ thick] (-1,0) -- (0,-2) -- (1,0);
    \draw[ thick] (-1,0) -- (0,1) -- (1,0);
    \draw[ thick] (-1,0) -- (0,2) -- (1,0);
    \draw (0.0,2.16) node {\footnotesize{$m^{\prime}$}};
    \draw (0.16,1.04) node {\footnotesize{$u_3$}};
    \draw (0.16,0.12) node {\footnotesize{$u_2$}};
    \draw (0.16,-1.1) node {\footnotesize{$u_1$}};
\end{tikzpicture}
};
\node (web2) at (4.5,0){
\begin{tikzpicture}[scale=1.5]
    \draw[ thick] (0,0) -- (0,-2);
    \draw[ thick] (-1,0) -- (1,0);
    \draw[ thick] (-1,0) -- (0,-1) -- (1,0);
    \draw[ thick] (-1,0) -- (0,-2) -- (1,0);
    \draw (0,0.16) node {\footnotesize{$m$}};
    \draw (0.16,-1.1) node {\footnotesize{$u$}};
\end{tikzpicture}
};
\draw (-3.9,-3.9) node {(a)};
\draw (4.5,-3.9) node {(b)};
\end{tikzpicture}
\end{center}
\caption{(a) Toric diagram for the pure $SU(4)$ theory. (b) Toric diagram for the pure $SU(2)$ theory.}
\label{fig:embedding0}
\end{figure}
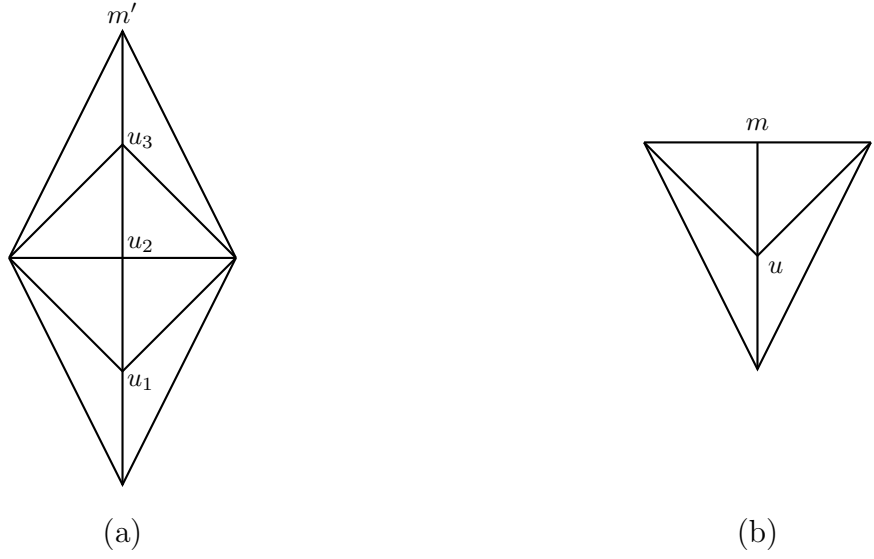

An illuminating example is the embedding of the geometry for the pure $SU(2)$ theory into the geometry for the pure $SU(4)$ theory as illustrated in Figure~\ref{fig:embedding0}. The geometry for the $SU(4)$ theory contains three compact surface components $D_1=\mathbb{F}_2$, $D_2=\mathbb{F}_0$ and $D_3=\mathbb{F}_2$ that corresponds to $u_1,u_2$ and $u_3$ respectively. By taking the volumes of $D_2$ and $D_3$ to infinity, we obtain the geometry depicted in Figure~\ref{fig:embedding0}(b) for the pure $SU(2)$ theory. The parameter $u_2$ corresponds to a non-compact surface and it becomes the mass parameter $m$ for the $SU(2)$ theory.

More precisely, for the pure $SU(2)$ theory given by Figure~\ref{fig:embedding0}(b), we have:
\begin{equation}
    z_1 = \frac{m}{u^2},\ z_2 = \frac{1}{m^2}.
\end{equation}
It is easy to see that $\mathcal{G}_{u}=\mathbb{Z}_2$ is generated by $g: (u,m) \mapsto (-u,m)$ and $ \mathcal{G}_{um} = \mathbb{Z}_4$ given by $g: (u,m) \mapsto (iu,-m)$. For the pure $SU(4)$ theory given by Figure~\ref{fig:embedding0}(a), we have:
\begin{equation}
    z_1 =\frac{u_2}{u_1^2} ,\ z_2 =\frac{1}{u_2^2},\ z_3 = \frac{u_1u_3}{u_2^2},\ z_4=\frac{m^{\prime}u_2}{u_3^2}.
\end{equation}
It is easy to see that $\mathcal{G}_{u'} \cong \mathbb{Z}_4$ which is generated by $(u_1,u_2,u_3,m^{\prime})\mapsto (iu_1,-u_2,-iu_3,m^{\prime})$. It is now obvious that $|\mathcal{G}_{u'}| \geq |\mathcal{G}_{u}|$ as expected, and the action of $\mathcal{G}_{um}$ is inherited from the action of $\mathcal{G}_{u^{\prime}}$.

\paragraph{Matching the 4D rank-1 KK theories}

It is also interesting to match our result to the well-studied 4D rank-1 KK theories~\cite{Closset:2021lhd}. For a rank-1 4D KK theory obtained from the circle compactification of a 5D $\mathcal{N}=1$ theory, electically charged particles on the generic point of the Coulomb branch moduli space become dyons that are charged under $U(1)_q\times U(1)_m \times \mathbb{Z}(G_F)$ where $Z(G_F) = \mathbb{Z}_{n_1} \times \cdots \times \mathbb{Z}_{n_p}$ and $G_F$ is the simply-connected group associated to the flavor symmetry algebra $\mathfrak{g}_F$. Thus the charge of a dyon state $\psi$ is:
\begin{equation}
	(q, m, l_1, \cdots, l_p),\ l_i \in \mathbb{Z}_{n_i}.
\end{equation}
We define $\mathcal{E}\subset U(1)_q\times U(1)_m \times \mathbb{Z}(G_F)$ as the group of transformations leaves the dyon spectrum invariant. Denote the generators of $\mathcal{E}$ by:
\begin{equation}\label{eq:gE_rank1}
	g^{\mathcal{E}}: (k_q, k_m, t_1, \cdots, t_p),\ k_{q,m}\in \mathbb{Q},\ t_i\in \mathbb{Z}_{n_i},
\end{equation}
for a dyon $\psi$ we have:
\begin{equation}\label{eq:rank1_dyon_trans}
	g^{\mathcal{E}}: \psi \mapsto \exp\left( 2\pi i \left( k_q q + k_m m + \sum_{i=1}^p \frac{t_i l_i}{n_i}  \right) \right) \psi.
\end{equation}
We define $\mathcal{Z}^{[1]}$ to be group generated by:
\begin{equation}
    g^{\mathcal{Z}^{[1]}} = (k_1, k_m, 0, \cdots, 0).
\end{equation}
One can thus form the short exact sequence:
\begin{equation}\label{eq:Closset_SES}
    0 \rightarrow \mathcal{Z}^{[1]} \rightarrow \mathcal{E} \rightarrow \mathcal{F} \rightarrow 0
\end{equation}
where $\mathcal{Z}^{[1]}\cong \Gamma^{(1)}$ is the 1-form symmetry group of the rank-1 theory and the SES itself will be the 2-group symmetry of the theory if it does not split. One can immediately observe that~(\ref{eq:Closset_SES}) shares essentially the same form as~(\ref{eq:Extend_Gum_Gu}), expect for the direction of the arrows which is unimportant for our purpose.

This is not a coincidence if we take a closer look at~(\ref{eq:rank1_dyon_trans}). A charged state is given by an M2-brane wrapping mode on certain 2-cycle and its charge is given by the intersection numbers of that 2-cycle with various 4-cycles in the geometry. In other words, the charges are given by $Q^G$ and $Q^F$ in~\eqref{eq:z_um}. We focus on the M2-brane wrapping modes where the charges are $(q,l_1,\cdots,l_p)$ and~\eqref{eq:z_um} becomes:
\begin{equation}
    z_i = u^{q_i} \prod_p m_p^{l_{ip}}.
\end{equation}
Comparing the above equation with~(\ref{eq:rank1_dyon_trans}), it is not hard to show that $\mathcal{G}_{um} \cong \mathcal{E}$ for $\mathcal{G}_{um}$ defined in Section~\ref{sec:main_result} and $\mathcal{E}$ generated by~(\ref{eq:gE_rank1}). Therefore we see that our approach nicely reproduces the 1-form and 2-group (when~(\ref{eq:Closset_SES}) does not split) symmetries of rank-1 theories as discussed in e.g.~\cite{Morrison:5D_higher_form, Bhardwaj:Higher_form_5D6D, Closset:2021lhd, Apruzzi:GlobalForm_2group, Hubner:GenSymm_EFib, Cvetic:0form_1form_2group}.

\subsection{Mordell-Weil group of special geometry and 2-group symmetry}\label{sec:conjecture}

For rank-1 KK theories the spectrum~(\ref{eq:gE_rank1}) is related to the Mordell-Weil group of the Coulomb branch (CB) geometry $E\hookrightarrow \mathcal{S} \xrightarrow{\pi} \mathbb{P}^1$ where $\mathbb{P}^1$ is the $U$-plane with the additional point at infinity and $E$ the Seiberg-Witten curve~\cite{Closset:Uplane}. More precisely, the SES~(\ref{eq:Closset_SES}) is matched with the following SES:
\begin{equation}
    0 \rightarrow \mathcal{Z}^{[1]} \rightarrow \Phi_{\text{tor}} \rightarrow \mathcal{F} \rightarrow 0
\end{equation}
where $\Phi_{\text{tor}}$ is the torsion subgroup of the Mordell-Weil group $\Phi$ of $\mathcal{S}$ when $\text{rank}(\Phi) = 0$, i.e. when $\Phi$ is purely torsional. In particular, it was proposed in~\cite{Closset:Uplane, Closset:2023pmc, Furrer:2024zzu} that $\Phi_{\text{tor}}$ is the 2-group symmetry of the corresponding theory on its CB~\footnote{Usually the 2-group symmetry is defined to be the whole short exact sequence. For simplicity, in this section we adopt the convention in~\cite{Closset:Uplane} that the middle element in the sequence is called the 2-group symmetry on the CB.}.

It is interesting to see if the above conjecture can be generalized to higher rank 4D KK theories, where $\mathcal{S}$ is replaced by higher dimensional CB geometries~\cite{Strominger:1990pd, Argyres:2020nrr}, i.e. the higher dimensional generalizations of the $U$-plane of the $S^1$-compactified 5D SCFT proposed in~\cite{Closset:Uplane}~\footnote{We emphasize that the 5D theory we consider in this work is actually in $\mathbb{R}^{4}\times S^1$.}. The developments in~\cite{Caorsi:2018ahl, Cecotti:2024jbt} will be useful for us to conjecture a similar relation between the 2-group symmetry of higher rank theories and certain geometric property of its CB.

Following~\cite{Cecotti:2024jbt}, for rank $r > 1$ the elliptic surface $\mathcal{S}$ is generalized to a fibration $\pi:\mathscr{Y}\rightarrow \mathscr{B}$ where the generic fiber $\mathscr{Y}_b$ for $b \in \mathscr{B}$ is an anti-affine group which is the extension of a complex $r$-dimensional Abelian variety $\mathscr{A}_b$ by an additive (vector) group $V \cong \mathbb{C}^k$ and a multiplicative (torus) group $T \cong (\mathbb{C}^*)^f$ and
\begin{equation}
    \mathscr{B} = \mathscr{C} \times \mathscr{P}
\end{equation}
where $\mathscr{C}$ is the ordinary CB parameterized by the parameters $u_1,\dots,u_r$ and $\mathscr{P}$ the space of the mass parameters $m_1,\dots,m_f$. The total space of the fibration $\varpi:\mathscr{A} \rightarrow \mathscr{B}$ with generic fiber $\mathscr{A}_b$ is the \emph{universal special geometry} of the $\mathcal{N} = 2$ theory which can be viewed as the family of ordinary special geometries parameterized by $\mathscr{P}$. The ordinary special geometry at a fixed coupling $p \in \mathscr{P}$ is $\mathscr{X} := \mathscr{A}|_{p}$. The Mordell-Weil group $MW(\mathscr{A})$ is the group of global sections of $\varpi :\mathscr{A} \rightarrow \mathscr{B}$~\cite{schütt2019mordell}. For $MW(\mathscr{A})$ to be finitely generated, we require the Chow trace of $\mathscr{A}$ be zero~\cite{chow1955abelian, Lang&Neron}.

By restriction to $p\in\mathscr{P}$, a section of the universal special geometry $\mathscr{A}\rightarrow \mathscr{B}$ defines a section of the ordinary special geometry $\mathscr{X}\rightarrow \mathscr{C}$, i.e. we have:
\begin{equation}
    MW(\mathscr{A}) \subset MW(\mathscr{X}).
\end{equation}
There is a pair $(M,S)$ with $M$ projective such that $\mathscr{X} = M \setminus S$ for a divisor $S \subset M$. By Oguiso's generalization of Shioda-Tate formula~\cite{schütt2019mordell, Oguiso2007SHIODATATEFF} one can calculate $\text{rank}(MW(\mathscr{X}))$ using the data of $M$ and $S$. E.g. when $r = 1$, $M \cong \mathcal{S}$ is an elliptic surface over $\mathbb{P}^1$ and $S$ is a fiber so that $M\setminus S$ is symplectic and the Shioda-Tate formula can be applied.

Following the recipe in~\cite{Closset:Uplane} we choose $p_0\in \mathscr{P}$ such that $\text{rank}(MW(\mathscr{X}_{p_0}))$ is minimized. For $r = 1$ theories, $\text{rank}(MW(\mathscr{X})) = 0$ when all mass deformations are turned off, i.e. $MW(\mathscr{X})$ is purely torsion in the massless limit. Similarly when $r > 1$ we would expect that $\text{rank}(MW(\mathscr{X}_{p_0}))$ is minimized when all relevant deformations are turned off. We define:
\begin{equation}
    \Phi_{\text{tor}} = \text{Tor}(MW(\mathscr{X}_{p_0})).
\end{equation}
We \emph{conjecture} that $\Phi_{\text{tor}}$ is the 2-group symmetry of the theory corresponding to the ordinary special geometry $\mathscr{X}\rightarrow\mathscr{C}$. When $r = 1$ this reduces to the result in~\cite{Closset:Uplane} when $\mathcal{X} \equiv \mathcal{S}$ is \emph{extremal}.

To further extract 1-form symmetry $\mathcal{Z}^{[1]}$ from the geometry of $\mathscr{A}$ (or $\mathscr{X}$) we shall consider its resolution $\widetilde{\mathscr{A}}$. It is expected that for some $\mathcal{N}=2$ QFT certain singularities at codimension $\geq 2$ of $\mathscr{A}$ has no crepant resolution~\cite{Cecotti:2024jbt} but we will not study these subtleties in this work. Nevertheless we will assume that the codimension-1 singularities of $\mathscr{A}$ can be resolved crepantly, i.e. we resolve the non-smooth fiber along the \emph{discriminant locus} $\mathscr{D} \subset \mathscr{B}$ while keeping $\mathscr{A}$ symplectic. The connected irreducible component $\mathscr{A}_b^\circ \subset \mathscr{A}_b$ that intersects the \emph{zero section} $\pi^{-1}(\mathscr{B})$ is thus the ``affine node'' of $\widetilde{\mathscr{A}}_b$. We then define the \emph{narrow sections} in a similar fashion to that in~\cite{Closset:Uplane}:
\begin{equation}
    \Phi_{\text{narrow}} = \{ P\in\Phi_{\text{tor}} | P\cdot \mathscr{A}_b^\circ \neq 0,\ b \in \mathscr{D}\ \text{and}\ b \notin \pi(S) \}
\end{equation}
where the restriction from $\mathscr{A}$ to $\mathscr{X}$ is assumed. We \emph{conjecture} that $\Phi_{\text{narrow}}$ is the 1-form symmetry of the theory.

The above conjectures are motivated by the conjecture in~\cite{Closset:Uplane} to which they reduce almost trivially when $r = 1$. We point out that these conjectures can potentially be verified by studying carefully the charged spectrum as was done in rank-1 case in~(\ref{eq:gE_rank1}) since the root lattice of the flavor group has a nice embedding into $\mathscr{A}_b^\vee$. Furthermore, the 2-group symmetries of these higher rank cases can certainly be cross-checked against our B-model approach purposed in Section~\ref{sec:main_result}, since after all the complex structure moduli can be viewed as functions on at least locally on $\mathscr{B}$, e.g. they can be written explicitly as~(\ref{eq:diagnal_ziui}). Another interesting issue is to relate the Mordell-Weil torsion of $\mathscr{X}$ with the monodromy in the complex structure moduli space as heuristically outlined in Section~\ref{sec:motivation} and further as is concretely exemplified by~(\ref{eq:diagnal_ziui}). We will leave a more detailed discussion of these issues in future works.

\section{Conclusion and discussion}\label{sec:conclusion}
In this work, we show that the 1-form and the 2-group symmetry of a 5D SCFT $\mathcal{T}_{M_6}$ can be related to the monodromy group at large radius point for the B-model via mirror symmetry. To be more precise, we start by constructing the half-BPS Wilson line defects as M2-brane wrapping (torsional) relative 2-cycles in a non-compact CY3 $X_6$. Via mirror symmetry, the partition function as well as the VEVs for the line defects are expressed as functions of the complex structure parameters $z_j$ for the mirror manifold $W_6$. Using the B-model parameters $u_i$ and $m_i$, which correspond to the compact and non-compact surfaces of the Calabi-Yau 3-fold $M_3$, the complex structure parameter $z_j$ is expressed as a functional form of $z_j = z_j(u_i,m_i)$. The group $\mathcal{G}_u$ and $\mathcal{G}_{um}$, which is defined from \eqref{eq:trans} leaving the invariance of $z_j$ or at most up to a phase $z_j\mapsto e^{2\pi i} z_j$, determine the 2-group symmetry of the 5D SCFT. More precisely, the 2-group symmetry of the 5D SCFT:
\begin{equation}\label{eq:2group_SES_conclusion}
    0 \rightarrow \mathcal{F} \rightarrow \mathcal{E} \rightarrow \Gamma^{(1)} \rightarrow 0
\end{equation}
is isomorphic to the SES:
\begin{equation}
    0 \rightarrow \mathcal{F} \rightarrow \mathcal{G}_{um} \rightarrow \mathcal{G}_{u} \rightarrow 0.
\end{equation}

We also calculate VEVs of Wilson loop operators for various examples from mirror symmetry. Those include 5D pure gauge theories with exceptional gauge groups, where the calculations can not be performed using the localization method. By first constructing the geometry from compact Calabi-Yau hypersurface and then using Picard-Fuchs operators, we successfully obtained recursive relations for the instanton contribution for VEVs of Wilson loops.  

One advantage of our discussion is that it can be readily applied to the study of 5D supergravity theories arising from M-theory compactifications on compact Calabi-Yau threefolds. For example, in a recent work \cite{Huang:2025xkc}, the authors compute the refined BPS numbers of such threefolds and demonstrate that a one-form symmetry naturally emerges in the weak gravity limit due to B-field shifts. This symmetry, in turn, constrains the Wilson loop expansion of the refined BPS partition functions in 5D supergravity theories.

It is interesting to further explore the relation between various terms in the 2-group symmetry exact sequence~(\ref{eq:2group_SES_conclusion}) and the special geometry of the theory as discussed in Section~\ref{sec:phys_consequences}. In this direction it will be interesting to look at the so-called \emph{torsion conjecture} on the bound of the torsion group of abelian varieties, of which the simplest case is Mazur's theorem on the torsion group of the elliptic curves~\cite{Mazur1977,Mazur1978}~\footnote{We thank Min-xin Huang for mentioning this theorem to us and inspiring us to look further into this series of conjectures.}. Based on our discussion in Section~\ref{sec:phys_consequences}, the torsion conjecture not only puts bounds on the Mordell-Weil group of the special geometry of the theory, it also greatly constrains the possible 2-group symmetries of a theory with 8 supercharges in 4D (as a KK theory) or in 5D. Therefore, it will be fruitful to further investigate these interesting relations along these lines.

The discussion in our work are straightforward to generalize to the study of higher-than-1 form symmetries, and it is interesting to explore it in the near future. Another exciting direction worth digging is to see how non-invertible symmetries and the VEVs of the corresponding symmetry operators can be constructed from geometric engineering perspective. Since in this work we studied only the leading order behavior of the relation between the complex structure moduli and the K\"ahler moduli via mirror symmetry (e.g.~(\ref{eq:z_um}) or~(\ref{eq:diagnal_ziui})), we expect that those more complicated symmetry properties, e.g. non-invertible symmetries, shall appear within higher order terms. We will leave the analysis of these higher order corrections in future work.

\vspace{.5cm}
\noindent \textbf{Acknowledgments.}
We thank Lakshya Bhardwaj, Mirjam Cveti\v{c}, James Halverson, Max H\"ubner, Ziming Ji, Albrecht Klemm, Kimyeong Lee, Jian-Xin Lu, Paul-Konstantin Oehlmann, Leonardo Santilli, Benjamin Sung, and Yi Zhang for helpful discussions. We would like to thank Min-xin Huang and Yi-Nan Wang for their invaluable comments on the draft. We thank the organizers and the participants of the SymTFT workshop at Peng Huanwu Center for Fundamental theory and of the Tianfu Fields and Strings 2024 conference at Chengdu where the work was presented for their comments and suggestions. JT would like to thank Ying Zhang for her love and support. JT is supported by National Natural Science Foundation of China under Grant No. 12405085 and by the Natural Science Foundation of Shanghai (Grant No. 24ZR1419300). XW is supported by the National Natural Science Foundation of China (Grants No.12247103).

\appendix

\section{An alternative approach to arrive at $\langle W \rangle$}\label{app:SNF_approach}

There is an alternative approach to arrive at the key equation~(\ref{eq:SUn_Wilson}) following the recipe in~\cite{Morrison:5D_higher_form, Bhardwaj:Higher_form_5D6D, Hubner:GenSymm_EFib}. We will give the derivation following this approach in this appendix.

We pick up from the definition of $\kappa_R$ around~(\ref{eq:find_cpt_reps}). Written in terms of an intersection pairing in $M_6$, $\kappa_R$ can be calculated via finding the compact representative $\Sigma_c \in H_2({\tM}, \mathbb{Q})$ of $\Sigma$ as mentioned in Section~\ref{sec:WilsonLine_M2brane}. The difference is now we will follow the approach in~\cite{Bhardwaj:Higher_form_5D6D, Hubner:GenSymm_EFib}. More precisely, there exists $\Sigma_c = \sum_i c_i \Sigma_i$ for $\Sigma_i \in H_2({\tM},\mathbb{Z})$ and $c_i\in \mathbb{Q}/\mathbb{Z}$ such that its intersection pairing with any 4-cycle in $H_4({\tM},\mathbb{Z})$ is the same as $\Sigma$. We will determine the coefficients $c_i$.

To determine $c_i$ we look at the intersection pairing:
\begin{equation}
    \mathcal{M}_4: H_4({\tM}, \mathbb{Z}) \times H_2({\tM}, \mathbb{Z}) \rightarrow \mathbb{Z}
\end{equation}
viewed as a $b_4\times b_2$ matrix where $b_i$ is the $i^{\text{th}}$ Betti number of ${\tM}$, whose \emph{Smith normal form}:
\begin{equation}
    \text{SNF}(\mathcal{M}_4) = U\mathcal{M}_4V = \begin{pmatrix}
        \alpha_{1} & 0 & \cdots & 0 & \cdots &0 \\
        0 & \alpha_{2} & \cdots & 0 & \cdots & 0 \\
        \vdots & & \ddots & \vdots & & \vdots \\
        0 & \cdots &  & \alpha_{b_4} & \cdots & 0
    \end{pmatrix},\ \alpha_i \in \mathbb{Z}_+
\end{equation}
is crucial for our purpose. It has been shown in~\cite{Hubner:GenSymm_EFib} that the $i^{\text{th}}$ column of $V$, after being normalized by $\alpha_i$, determines the coefficients $c_j^{(i)}$ mod 1 for $\Sigma^{(i)}\in H_2(M_6,\partial M_6)$. In other words, the $i^{\text{th}}$ column of $V$ is given by $\mathfrak{c}_j^{(i)} := \alpha_i c_j^{(i)} \mod \alpha_i$  , $j = 1,\cdots,b_2$.

On the other hand, the $i^{\text{th}}$ row of $U$, after being normalized by $\alpha_i$, determines the coefficients $\mathfrak{d}_j^{(i)}$ mod 1 for the \emph{center divisor}~\cite{Hubner:GenSymm_EFib}:
\begin{equation}
    \mathfrak{D}^{(i)} = \sum_j \mathfrak{d}_j^{(i)} D_j \in H_4(M_6, \mathbb{Q}).
\end{equation}
Given $\mathfrak{D}^{(i)}$, for the $i^{\text{th}}$ dual divisor we have $D^{(i)}_\Sigma = \mathfrak{D}^{(i)}/\alpha_i$. Note that here we have made use of a non-trivial fact that the Poincar\'e-Lefschetz dual of $\omega^\Sigma \in H^2(M_6,\partial M_6)$ is a center divisor due to the existence of the following well-defined, non-trivial pairing mod 1~\cite{Hubner:GenSymm_EFib}:
\begin{equation}
    H_4(M_6, \mathbb{Q}/\mathbb{Z}) \times H_2(M_6, \mathbb{Q}/\mathbb{Z}) \rightarrow \mathbb{Q}/\mathbb{Z},
\end{equation}
meaning that for each compact representative $\Sigma_c\in H_2(M_6, \mathbb{Q}/\mathbb{Z})$ of $\Sigma\in H_2(M_6, \partial M_6)$ there is a dual 4-cycle in $H_4(M_6, \mathbb{Q}/\mathbb{Z})$ which by definition is a center divisor.

Making use of $U$ and $V$, in particular the properties of coefficients $\mathfrak{c}_j^{(i)}$ and $\mathfrak{d}_j^{(i)}$, we arrive at the following result for the desired intersection pairings:
\begin{equation}\label{eq:desired_pairing}
    D_\Sigma^{(i)}\cdot \Sigma^{(j)} = \left( \sum_j \frac{\mathfrak{d}_j^{(i)}}{\alpha_i} D_j \right) \cdot \left( \sum_k \frac{\mathfrak{c}_k^{(j)}}{\alpha_i} \Sigma_k \right) = \begin{pmatrix}
        1/\alpha_{1} & 0 & \cdots & 0 & \cdots &0 \\
        0 & 1/\alpha_{2} & \cdots & 0 & \cdots & 0 \\
        \vdots & & \ddots & \vdots & & \vdots \\
        0 & \cdots &  & 1/\alpha_{b_4} & \cdots & 0
    \end{pmatrix}.
\end{equation}
Therefore we have $\kappa_R^{ij} = \frac{t^i}{\alpha_i}\delta^{ij} $.

As a consequence of~(\ref{eq:desired_pairing}), the VEV of a line defect operator~(\ref{eq:WilsonVEV}) becomes:
\begin{equation}\label{eq:WilsonVEV_multi}
    \langle W(\ell) \rangle = \prod_j \left\langle\ e^{\frac{2\pi i t^j}{\alpha_j} \int_\ell A_{j}} \cdots \right\rangle.
\end{equation}
We now ask the physical question of what the field $A_j$ is. It is clear that $A_j$ is obtained from reducing $C_3$ on the 2-form $\omega^\Sigma_{(j)}$. Hence, as mentioned at the beginning of this section, it is the gauge field corresponding to the center divisor $D_\Sigma^{(j)}$. Without loss of generality we assume there is only one center divisor $D_\Sigma$ therefore there is also only one gauge field $A$ corresponding to $D_\Sigma$. In the singular limit where the geometry becomes the blow-down $M_6$ of ${\tM}$, the gauge symmetry lifts to certain non-abelian $G$ and $A$ will be the $U(1)$-valued 1-form gauge field of the $Z(G)$-symmetry, where $Z(G)$ is the center of $G$.

For simplicity we further assume that $G = SU(n)$, in which case we have:
\begin{equation}
    \langle W(\ell) \rangle = \left\langle\ e^{\frac{2\pi i t}{n} \int_\ell A} \cdots \right\rangle.
\end{equation}
In the $R\rightarrow \infty$ limit, the dynamics of $A$ decouples hence $e^{\frac{2\pi i t}{n} \int_\ell A}$ factors out from the VEV. We have arrive at~(\ref{eq:SUn_Wilson}) as promised at the beginning of this section.

\section{Geometries for compact CY3s}
In this appendix, we construct compact elliptically fibered Calabi-Yau threefolds with singular fibers that are relevant to the discussion in Section \ref{sec:examples}. The compact geometry is a generic (singular) CY3 hypersurface $X$ in a toric ambient space $A_\Delta$ associated with a 4D reflexive polytope $\Delta$ and the method of the construction is standard \cite{batyrev1993dual}. We perform a resolution $\widetilde{X}\rightarrow X$ of $X$ by adding rays to $\Delta$. To obtain its non-compact limit, we choose suitable divisors of $\widetilde{X}$ and send its volume to infinity while keeping the volume of the other divisors finite by tuning the K\"ahler moduli of $\widetilde{X}$ using the same method employed in \cite{Halverson:2022mtc}. We then employ the method in \cite{Haghighat_2015, Del_Zotto_2018} to calculate the topological string partition function of these non-compact Calabi-Yau threefolds.

In the following, we present constructions of two relevant compact Calabi-Yau threefolds.

\subsection{$G_2+\mathbf{7}$ geometry}\label{sec:geom_G2}

We realize $G_2+\mathbf{7}$ model via decompactification of a compact model with $G_2+\mathbf{7}$ spectrum. Let us consider the reflexive polytope $\Delta$ which contains the following rays:
\begin{equation}\label{eq:geom_G2}
    \begin{split}
        &r_x = (-1,0,0,0),\ r_y = (0,-1,0,0),\ r_z = (2,3,0,0),\ r_u = (2,3,1,0),\ r_v = (2,3,0,-1), \\
        &r_s = (2,3,0,-2),\ r_t = (1,1,0,-1),\ r_p = (2,3,-1,-3),\ r_q = (1,2,-1,-3),\ r_{d} =(2,3,0,1).
    \end{split}
\end{equation}
It is standard to compute the equation of the compact CY3 hypersurface $X$ and show that there is a $G_2$ supported on the divisor $D_v \cap X$ via checking the corresponding monodromy cover \cite{Bershadsky_1996}. It is also easy to see that $D_v\cdot_X D_v\cdot_X D_z = -3$. Therefore by anomaly cancellation there is one hypermultiplet in $\mathbf{7}$ of $G_2$  \cite{Johnson_2016}.

\subsection{$SO(9)+\mathbf{9}_v$ geometry}

We realize $SO(9)+\mathbf{9}$ model via decompactification of a compact model with $SO(9)+\mathbf{9}$ spectrum. In this case let us consider the reflexive polytope $\Delta$ which contains the following rays:
\begin{equation}\label{eq:raysSO9}
    \begin{split}
        &r_x = (-1,0,0,0),\ r_y = (0,-1,0,0),\ r_z = (2,3,0,0),\ r_u = (2,3,1,0),\\
        &r_v = (2,3,0,-1),\ r_s = (2,3,-1,-4),\ r_g = (1,1,0,-1),\ r_t = (0,1,0,-1),\\
        &r_p = (1,2,0,-2),\ r_q = (2,3,0,-2),\ r_{d} =(2,3,0,1),\ r_{f}=(1,2,1,0).
    \end{split}
\end{equation}
Again it is standard to show that there is an $SO(9)$ group supported on $D_v\cap X$ where $X$ is the CY3 hypersurface and $D_v\cdot_X D_v\cdot_X D_z = -4$ therefore by anomaly cancellation there is one hypermultiplet in $\mathbf{9}_v$ of $SO(9)$  \cite{Johnson_2016}.

\section{5D $F_4$ theory}
\begin{equation} \label{F4polytope}
 \begin{array}{crrrr|rrrrrrr|} 
    &\multicolumn{4}{c}{\nu_i^*}     &l^{(1)} & l^{(2)} & l^{(3)} & l^{(4)} & l^{(b)} &  \\   
    D_{1}& 0 & 0 & 0 & 0 & 0 & -2 & 0 & 0 & 0 & \\
D_{2}& -1 & 0 & 0 & 0 & 0 & 0 & 0 & 1 & 0 & \\
D_{3}& 0 & -1 & 0 & 0 & 0 & 1 & 0 & 0 & 0 & \\
D_{4}& 2 & 3 & 0 & -1 & 1 & 0 & 0 & 0 & 0 & \\
D_{u_1}& 2 & 3 & 0 & -2 & -2 & 1 & 0 & 0 & -1 & \\
D_{u_2}& 2 & 3 & 0 & -3 & 1 & -2 & 1 & 0 & -1 & \\
D_{u_3}& 1 & 2 & 0 & -2 & 0 & 2 & -2 & 1 & 0 & \\
D_{u_4}& 0 & 1 & 0 & -1 & 0 & 0 & 1 & -2 & 0 & \\
D_{5}& 2 & 3 & -1 & -5 & 0 & 0 & 0 & 0 & 1 & \\
D_{6}& 2 & 3 & 1 & 0 & 0 & 0 & 0 & 0 & 1 & \\
\end{array} \  
\end{equation} 
The geometry of the 5D pure $F_4$ gauge theory is a hypersurface in the toric variety whose toric data is in equation \eqref{F4polytope}. This geometry can be obtained as non-compact limit for the compact geometry \cite[(4.22)]{Haghighat:2014vxa}. The invariant coordinates are
\begin{equation}\label{eq:F4zu1}
    z_1=\frac{u_2}{u_1^2},\quad  z_2=\frac{u_1u_3^2}{u_2^2},\quad z_3=\frac{u_2u_4}{u_3^2},\quad z_4=\frac{u_3}{u_4^2},\quad z_5=\frac{m}{u_1u_2},
\end{equation}
where $m=\frak{q}$ is related to the non-compact surface $D_5$ or $D_6$. Under this parameterization, the parameter $m$ is the instanton counting parameter for the 5D pure $F_4$ theory. One can observe the groups $\mathcal{G}_{u}$ and $\mathcal{G}_{um}$ are trivial. To compute the Wilson loops, we first compute the mirror maps via the ansatz
\begin{align}
    t_i=\log(z_i)+\mathcal{O}(z)=f_{0,i}(u_1,u_2,u_3,u_4)+\sum_{n=1}^{\infty}f_{n,i}(u_1,u_2,u_3,u_4)m^n.
\end{align}
The leading term $f_{0,i}$ can be easily solved via Picard-Fuchs equations and the assumption that the $u_i$ are written exactly via $F_4$ characters. The instanton corrections of $u_i$'s can be solved via the Picard-Fuchs operator:
\begin{align}
    \mathcal{L}=\theta _5^2+z_5\left(\theta _1-2 \theta _2+\theta _3-\theta _5\right) \left(2 \theta
   _1-\theta _2+\theta _5\right) =m\partial_m+m^2\partial_m^2-m\partial_{u_1}\partial_{u_2},
\end{align}
we obtain the recursion relations 
\begin{align}
    f_{n,i}(u_1,u_2,u_3,u_4)=\frac{1}{n^2}\partial_{u_1}\partial_{u_2}f_{n-1,i}(u_1,u_2,u_3,u_4),\quad n>0.
\end{align}
At 0- and 1-instanton level, we find
\begin{align}
    u_1&=\chi_{\mathbf{52}}-3\chi_{\mathbf{26}}+26+\frac{\mathcal{N}_1}{\mathcal{D}}\frak{q}+\mathcal{O}(\frak{q}^2),\\
    u_2&=2 \chi_{\mathbf{26}}^2-\chi_{\mathbf{52}} \chi_{\mathbf{26}}+50 \chi_{\mathbf{26}}+20 \chi_{\mathbf{52}}+\chi_{\mathbf{1274}}-11 \chi_{\mathbf{273}}-611+\frac{\mathcal{N}_2}{\mathcal{D}}\frak{q}+\mathcal{O}(\frak{q}^2),\\
    u_3&=-2 \chi_{\mathbf{52}}+\chi_{\mathbf{273}}-15 \chi_{\mathbf{26}}+221+\frac{\mathcal{N}_3}{\mathcal{D}}\frak{q}+\mathcal{O}(\frak{q}^2),\\
    u_4&=\chi_{\mathbf{26}}-26+\frac{\mathcal{N}_4}{\mathcal{D}}\frak{q}+\mathcal{O}(\frak{q}^2).
\end{align}
where $\chi_{\mathbf{52}},\chi_{\mathbf{1274}},\chi_{\mathbf{273}}$ and $\chi_{\mathbf{26}}$ are characters for the representations $\mathbf{52},\mathbf{1274},\mathbf{273}$ and $\mathbf{26}$ respectively, $\mathcal{D}$ is the product over longroots for $F_4$ and $\mathcal{N}_i$ are the numerators. With the help of the package LieART 2.0 \cite{Feger:2019tvk}, we manage to express the numerators in terms with the characthers $\chi_{\mathbf{R}}$ of $F_4$:
\begin{align*}
    \mathcal{N}_1=\,\,&64 \chi_{\irrep{1}}-88 \chi_{\irrep{26}}+16 \chi_{\irrep{273}}-160 \chi_{\irrep{324}}+114 \chi_{\irrep{1274}}-20 \chi_{\irrep[1]{1053}}+20 \chi_{\irrep{1053}}-80 \chi_{\irrep{2652}}\\
    &+52 \chi_{\irrep{8424}}-24 \chi_{\irrep{10829}}+4 \chi_{\irrep{19448}}+12 \chi_{\irrep{29172}}-24 \chi_{\irrep{12376}}+46 \chi_{\irrep{19278}}-44 \chi_{\irrep{34749}}\\
    &+8 \chi_{\irrep{16302}}-40 \chi_{\irrep{17901}}+16 \chi_{\irrep{76076}}+14 \chi_{\irrep{107406}}-12 \chi_{\irrep{119119}}+4 \chi_{\irrep{160056}}+8 \chi_{\irrep[1]{160056}}\\
    &-12 \chi_{\irrep{205751}}-4 \chi_{\irrep{420147}}+2 \chi_{\irrep{226746}}+4 \chi_{\irrep{379848}},
\end{align*}
\begin{align*}    
    \mathcal{N}_2=\,\,&6412 \chi_{\irrep{1}}+8000 \chi_{\irrep{26}}+1032 \chi_{\irrep{52}}-1522 \chi_{\irrep{273}}+3780 \chi_{\irrep{324}}-3360 \chi_{\irrep{1274}}-458 \chi_{\irrep[1]{1053}}\\
    &-1248 \chi_{\irrep{1053}}-224 \chi_{\irrep{4096}}+2564 \chi_{\irrep{2652}}-1480 \chi_{\irrep{8424}}+1948 \chi_{\irrep{10829}}-664 \chi_{\irrep{19448}}\\
    &-288 \chi_{\irrep{29172}}+120 \chi_{\irrep{12376}}-600 \chi_{\irrep{19278}}+582 \chi_{\irrep{34749}}-652 \chi_{\irrep{16302}}+646 \chi_{\irrep{17901}}\\
    &+96 \chi_{\irrep{106496}}-156 \chi_{\irrep{76076}}+92 \chi_{\irrep{160056}}-112 \chi_{\irrep{212992}}+52 \chi_{\irrep{81081}}+86 \chi_{\irrep{205751}}\\
    &-64 \chi_{\irrep{420147}}+2 \chi_{\irrep{340119}}+4 \chi_{\irrep{100776}}-144 \chi_{\irrep{107406}}-14 \chi_{\irrep{119119}}+52 \chi_{\irrep[1]{160056}}\\
    &+20 \chi_{\irrep{379848}}+20 \chi_{\irrep{184756}}+24 \chi_{\irrep{226746}}-16 \chi_{\irrep{787644}}+20 \chi_{\irrep{412776}}+32 \chi_{\irrep{1118208}}\\
    &+12 \chi_{\irrep{629356}}+4 \chi_{\irrep{952952}}-16 \chi_{\irrep{1327104}}-4 \chi_{\irrep{1002456}}-4 \chi_{\irrep{1801371}}+2 \chi_{\irrep{1074944}}\\
    &-12 \chi_{\irrep{1341522}}+4 \chi_{\irrep{2792556}}-4 \chi_{\irrep{1484406}}+4 \chi_{\irrep{3921372}}-2 \chi_{\irrep{2488563}}-4 \chi_{\irrep{3508596}}\\
    &+2 \chi_{\irrep{4582656}},
\end{align*}
\begin{align*}  
    \mathcal{N}_3=\,\,&-1152 \chi_{\irrep{1}}-1776 \chi_{\irrep{26}}-244 \chi_{\irrep{52}}+152 \chi_{\irrep{273}}-964 \chi_{\irrep{324}}+740 \chi_{\irrep{1274}}+104 \chi_{\irrep[1]{1053}}\\
    &+304 \chi_{\irrep{1053}}+28 \chi_{\irrep{4096}}-604 \chi_{\irrep{2652}}+380 \chi_{\irrep{8424}}-360 \chi_{\irrep{10829}}+148 \chi_{\irrep{19448}}+64 \chi_{\irrep{29172}}\\
    &-52 \chi_{\irrep{12376}}+172 \chi_{\irrep{19278}}-192 \chi_{\irrep{34749}}+80 \chi_{\irrep{16302}}-152 \chi_{\irrep{17901}}-12 \chi_{\irrep{106496}}+20 \chi_{\irrep{76076}}\\
    &+52 \chi_{\irrep{107406}}-16 \chi_{\irrep{119119}}-12 \chi_{\irrep{160056}}-4 \chi_{\irrep[1]{160056}}-8 \chi_{\irrep{205751}}+14 \chi_{\irrep{212992}}+8 \chi_{\irrep{420147}}\\
    &-4 \chi_{\irrep{226746}}-4 \chi_{\irrep{379848}}-4 \chi_{\irrep{1118208}}+2 \chi_{\irrep{1327104}},
\end{align*}
\begin{align*}  
    \mathcal{N}_4=\,\,&48 \chi_{\irrep{1}}+96 \chi_{\irrep{26}}+16 \chi_{\irrep{52}}+2 \chi_{\irrep{273}}+60 \chi_{\irrep{324}}-40 \chi_{\irrep{1274}}-6 \chi_{\irrep[1]{1053}}-16 \chi_{\irrep{1053}}+36 \chi_{\irrep{2652}}\\
    &-24 \chi_{\irrep{8424}}+16 \chi_{\irrep{10829}}-8 \chi_{\irrep{19448}}-4 \chi_{\irrep{29172}}+4 \chi_{\irrep{12376}}-12 \chi_{\irrep{19278}}+8 \chi_{\irrep{17901}}\\&+14 \chi_{\irrep{34749}}-4 \chi_{\irrep{107406}}+2 \chi_{\irrep{119119}}.
\end{align*}

\bibliographystyle{JHEP} 

\bibliography{refs}

\end{document}